\documentclass[12pt,aps,prd,preprint,tightenlines,superscriptaddress,
amsmath,amssymb,nofootinbib]{revtex4}

\RequirePackage[colorlinks=true
,urlcolor=blue
,anchorcolor=blue
,citecolor=blue
,filecolor=blue
,linkcolor=blue
,menucolor=blue
,pagecolor=blue
,linktocpage=true
,pdfproducer=medialab
,pdfa=true
]{hyperref}

\allowdisplaybreaks

\usepackage{amsmath,amssymb,amsthm,amsfonts}
\usepackage{graphicx}
\usepackage{subfigure}
\usepackage{dcolumn}	% Align table columns on decimal point
\usepackage{hyperref}      	% hypertext links %%ARXIV
\usepackage{bm}		% bold math
\usepackage{epsfig}
\usepackage{epstopdf}
\usepackage{setspace}
\usepackage[usenames, dvipsnames]{color}
\usepackage{slashed}
\usepackage{comment}
\usepackage{enumitem}
\usepackage[normalem]{ulem}
\usepackage[utf8]{inputenc}

\newcommand{\PRE}[1]{{#1}} % Use if preprint style

\newcommand{\be}{\begin{equation}\begin{aligned}}
\newcommand{\ee}{\end{aligned}\end{equation}}
\newcommand{\beq}{\begin{equation}}
\newcommand{\eeq}{\end{equation}}
\newcommand{\beqa}{\begin{eqnarray}}
\newcommand{\eeqa}{\end{eqnarray}}
\newcommand{\nn}{\nonumber}

\newcommand{\mev}{\text{MeV}}
\newcommand{\gev}{\text{GeV}}
\newcommand{\tev}{\text{TeV}}

\newcommand{\s}{\text{s}}

\renewcommand{\eqref}[1]{Eq.~(\ref{#1})}
\newcommand{\eqsref}[2]{Eqs.~(\ref{#1}) and (\ref{#2})}
\newcommand{\secref}[1]{Sec.~\ref{sec:#1}}

\newcommand{\figref}[1]{Fig.~\ref{fig:#1}}
\newcommand{\figsref}[2]{Figs.~\ref{fig:#1} and \ref{fig:#2}}

\newcommand{\appref}[1]{Appendix~\ref{sec:#1}}
\newcommand{\appsref}[2]{Appendixes~\ref{sec:#1} and \ref{sec:#2}}

\newcommand{\order}[1]{\mathcal{O}(#1)}

\def\be{\begin{equation}}
\def\ee{\end{equation}}
\def\bea{\begin{eqnarray}}
\def\eea{\end{eqnarray}}\def\nn{\nonumber}
\def\gsim{\ \rlap{\raise 2pt\hbox{$>$}}{\lower 2pt \hbox{$\sim$}}\ }
\def\lsim{\ \rlap{\raise 2pt\hbox{$<$}}{\lower 2pt \hbox{$\sim$}}\ }
\def\dslash{\kern-4pt \not{\hbox{\kern-2pt $\partial$}}}
\def\pslash{\not{\hbox{\kern-2pt p}}}

\def\gev{{\rm GeV }}
\def\l{{\rm L}}

\definecolor{gray}{rgb}{0.90,1,1}
\definecolor{LightCyan}{rgb}{0.88,1,1}

\def\bsmumu{b \to s \mu^+ \mu^-}

\def \s{\sqrt{2}}

\def \SM{{\rm SM}}

\def \NP{{\rm NP}}

\def \be{\beta}

\def \s{\sqrt{2}}

\def\beq{\begin{equation}}
\def\eeq{\end{equation}}
\def\bea{\begin{eqnarray}}
\def\eea{\end{eqnarray}}
\def\ber{\begin{eqnarray*}}
\def\eer{\end{eqnarray*}}
\def\bwt{\begin{widetext}}
\def\ewt{\end{widetext}}
\def\nn{\nonumber}
\def\roughly#1{\mathrel{\raise.3ex\hbox
{$#1$\kern-.75em\lower1ex\hbox{$\sim$}}}}
\def\lsim{\roughly<}
\def\gsim{\roughly>}

\def\order{\lower 1.8ex \hbox{\LARGE\~{}}}

\def\RK{R_K}
\def\RKstar{R_{K^*}}

\usepackage{bm}

\newcommand{\bctaunutau}{b \to c \tau^- {\bar\nu}_\tau}

\def \({\left(}
\def \){\right)}
\def \[{\left[}
\def \]{\right]}
\def \l|{\left|}
\def \r|{\right|}

\def \nn{\nonumber}

\def \be{\beta}

\def \SM{{\rm SM}}

\def \NP{{\rm NP}}

\def \s{\sqrt{2}}

\def\bsmumu{ b \to  s \mu^+ \mu^-}

\def\gev{{\rm GeV}}

\def\RK{R_K}

\def\RKstar{R_{K^*}}

\def \RD{R({D^{(*)}})}

\def \({\left(}
\def \){\right)}
\def \[{\left[}
\def \]{\right]}
\def \l|{\left|}
\def \r|{\right|}

\begin{document}

\preprint{UCI-TR-2019-10}
\preprint{UMISS-HEP-2019-02}

\title{\PRE{\vspace*{1.0in}}
Resolving the \boldmath{$(g-2)_{\mu}$} and \boldmath{$B$} anomalies \\
with leptoquarks and a dark Higgs boson
\PRE{\vspace*{.4in}}}

\author{Alakabha Datta}
\email{datta@phy.olemiss.edu}
\affiliation{Department of Physics and Astronomy, University of Mississippi, 108 Lewis Hall, Oxford, Mississippi 38677, USA
\PRE{\vspace*{.1in}}}
\affiliation{Department of Physics and Astronomy, University of California, Irvine, California 92697-4575, USA
\PRE{\vspace*{.1in}}}

\author{Jonathan L.~Feng}
\email{jlf@uci.edu}
\affiliation{Department of Physics and Astronomy, University of California, Irvine, California 92697-4575, USA
\PRE{\vspace*{.1in}}}

\author{Saeed Kamali}
\email{skamali@go.olemiss.edu}
\affiliation{Department of Physics and Astronomy, University of Mississippi, 108 Lewis Hall, Oxford, Mississippi 38677, USA
\PRE{\vspace*{.1in}}}
\affiliation{Department of Physics and Astronomy, University of California, Irvine, California 92697-4575, USA
\PRE{\vspace*{.1in}}}

\author{Jacky Kumar
\PRE{\vspace*{.5in}}}
\email{jacky.kumar@umontreal.ca}
\affiliation{Physique des Particules, Universite de Montreal, C.P. 6128, Succursale Centre-Ville,\\
Montreal, Qu\' ebec, Canada H3C 3J7
\PRE{\vspace*{.5in}}}

%\date{\today}

\begin{abstract}
\PRE{\vspace*{.2in}}

At present, there are outstanding discrepancies between standard model predictions and measurements of the muon's $g-2$ and several $B$-meson properties.  We resolve these anomalies by considering a two-Higgs-doublet model extended to include leptoquarks and a dark Higgs boson $S$.  The leptoquarks modify $B$-meson decays and also induce an $S \gamma \gamma$ coupling, which contributes to the muon's $g-2$ through a Barr-Zee diagram.  We show that, for TeV-scale leptoquarks and dark Higgs boson masses $m_{S} \sim 10-200~\mev$, a consistent resolution to all of the anomalies exists.  The model predicts interesting new decays, such as $B \to K^{(*)} e^+ e^-$, $B \to K^{(*)} \gamma \gamma$, $K \to \pi \gamma \gamma$, and $h \to \gamma \gamma \gamma \gamma$, with branching fractions not far below current bounds.  
\end{abstract}

%\pacs{}

\maketitle

%\tableofcontents

%\clearpage

%%%%%%%%%%%%%%%%%%%%%%%%%%%%%%%%%%%
\section{Introduction}
\label{sec:introduction}
%%%%%%%%%%%%%%%%%%%%%%%%%%%%%%%%%%%

At present, there are a number of anomalies in low-energy measurements.  Among these are the anomalous magnetic moment of the muon, $(g-2)_{\mu}$, and several in the decays of $B$ mesons.  Although none of these currently rises to the level of a $5 \, \sigma$ anomaly on its own, they are significant deviations, and it is interesting to investigate them, particularly if there are parsimonious explanations and if these explanations motivate new analyses of current and near-future data.   

In this work, we explain all of these anomalies in a concrete model: a two-Higgs-doublet model (2HDM) extended to include TeV-scale leptoquarks and a light scalar $S$ with mass $m_S \sim 10-200~\mev$.  We find solutions that depend on only a small number of parameters and show that these explanations motivate interesting new searches, particularly for rare meson decays to diphoton final states and Higgs boson decays to four photons.

The most longstanding anomaly we consider is in the anomalous magnetic moment of the muon.  A recent evaluation of the standard model (SM) prediction~\cite{Blum:2018mom} finds a $3.7 \, \sigma$ discrepancy with the experimental measurement~\cite{Bennett:2006fi}:
\begin{equation}
(g-2)_{\mu}^{\text{exp}} - (g-2)_{\mu}^{\text{SM}} = 27.4 \, (2.7) \, (2.6) \, (6.3) \times 10^{-10} \, .
\end{equation}
The first two uncertainties are theoretical and the last is experimental.  The experimental uncertainty is currently the largest, but it is expected to be reduced by a factor of 4 by the Muon $g-2$ Experiment~\cite{Grange:2015fou}, which is currently collecting data at Fermilab.  

In the $B$ sector, there are a large number of anomalies with various levels of significance; for a review, see Ref.~\cite{Kou:2018nap}. These anomalies may be divided into charged current (CC) processes, such as $ \bctaunutau$, and neutral current (NC) processes, such as $ b \to s \ell^+ \ell^-$.  The CC decays $B \to D^{(\ast)} \tau  \nu_\tau$  have been measured by the BABAR~\cite{BaBar1, BaBar2}, Belle~\cite{RD_Belle1, RD_Belle2, RD_Belle3}, and LHCb~\cite{RD_LHCb} Collaborations. These results may be expressed in terms of the ratios $\RD \equiv BR (\bar{B} \to D^{(*)} \tau^{-} {\bar\nu}_\tau)/ BR (\bar{B} \to D^{(*)} \ell^{-} {\bar\nu}_\ell)$, where $\ell = e,\mu$, in which many theoretical and systematic uncertainties cancel. By averaging the most recent measurements, the HFLAV Collaboration has found~\cite{HFAG}
\bea
R(D)^{\text{exp}} &=& 0.407 \pm 0.039 \pm 0.024  \\
R(D^{\ast})^{\text{exp}} &=& 0.304 \pm 0.013 \pm 0.007 \ ,
\label{ratiotau}
\eea
where, here and in the following, the first uncertainty is statistical and the second is systematic. These measurements exceed the SM predictions $R(D)^{\text{SM}} = 0.299 \pm 0.003$~\cite{Bigi:2016mdz} and $R(D^{\ast})^{\text{SM}} = 0.258 \pm 0.005$~\cite{Jaiswal:2017rve} by 2.3$\sigma$ and 3.4$\sigma$, respectively.  A combined analysis of $R(D)$ and $R(D^{\ast})$, including measurement correlations, finds a deviation of 4.1$\sigma$ from the SM prediction~\cite{HFAG}. A new measurement~\cite{ Abdesselam:2019dgh} by the Belle Collaboration, using semileptonic tagging, gives
\bea
R(D)^{\text{exp}} &=& 0.307 \pm 0.037 \pm 0.016  \\
R(D^{\ast})^{\text{exp}} &=& 0.283 \pm 0.018 \pm 0.014 \ ,
\label{ratiotau_new}
\eea
which reduces the deviation of the combined measurements from the SM predictions to about $3.1 \, \sigma$.

In the NC sector, the ratio $R_K \equiv BR(B^+ \to K^+ \mu^+ \mu^-)/BR(B^+ \to K^+ e^+ e^-)$~\cite{Hiller:2003js, Hiller:2014yaa} has been precisely measured by LHCb, most recently in Ref.~\cite{Aaij:2019wad}, which finds
\beq
R_K^{\text{exp}} = 0.846\, {}^{+0.060}_{-0.054}\, {}^{+ 0.016}_{-0.014} \, ,  \ 1 \le q^2 \le 6.0 ~{\rm GeV}^2 \ ,
%Old value: R_K^{\text{exp}} = 0.745\, {}^{+0.090}_{-0.074} \pm 0.036 \, ,  \ 1 \le q^2 \le 6.0 ~{\rm GeV}^2 \ ,
\label{RKexpt}
\eeq
where $q^2 = m_{\ell^+ \ell^-}^2$. This is lower than the SM prediction $R_K^{\text{SM}} = 1.00 \pm 0.01$~\cite{Bordone:2016gaq} by $2.5 \, \sigma$.  The related ratio $R_{K^*} \equiv BR(B^0 \to K^{*0} \mu^+ \mu^-)/BR(B^0 \to K^{*0} e^+ e^-)$ has been measured by LHCb to be~\cite{Aaij:2017vbb}\footnote{The Belle II Collaboration has also measured $R_{K^{(*)}}$~\cite{Abdesselam:2019wac,Abdesselam:2019lab} recently, but these measurements currently have relatively larger uncertainties and so have little effect on our analysis.}
\beq
R_{K^*}^{\text{exp}} = 
\left\{ 
\begin{array}{ll}
0.66\, {}^{+0.11}_{-0.07} \pm 0.03 \, , \quad & 0.045 \le q^2 \le 1.1~\gev^2 ~(\text{low} \ q^2) \\
0.69\, {}^{+0.11}_{-0.07} \pm 0.05 \, , \quad & 1.1 \le q^2 \le 6.0~\gev^2 ~ (\text{central}\ q^2) \ .
\end{array}
\right.
\eeq
These are also lower than the SM predictions~\cite{Bordone:2016gaq} $R_{K^{*}}^{\text{SM}} = 0.906 \pm 0.028$ (low $q^2$) and $R_{K^{*}}^{\text{SM}} = 1.00 \pm 0.01$ (central $q^2$) by $2.3 \, \sigma$ and $2.5 \, \sigma$, respectively.  Taken together, the general consensus is that these $B$-decay branching ratios differ significantly from SM predictions, and theoretical hadronic uncertainties~\cite{Descotes-Genon:2014uoa,Lyon:2014hpa,Jager:2014rwa} alone may not explain the data. 

An interesting question, then, is whether the $B$ anomalies have a common explanation in terms of new physics. Early work on the simultaneous explanation of the CC and NC anomalies~\cite{Bhattacharya:2014wla, Alonso:2015sja, Greljo:2015mma, Calibbi:2015kma} has been followed by many model calculations; an incomplete list can be found in Refs.~\cite {Altmannshofer:2017poe, Capdevila:2017bsm, Altmannshofer:2017yso, DAmico:2017mtc, Hiller:2017bzc, Geng:2017svp, Ciuchini:2017mik, Celis:2017doq, Becirevic:2017jtw, DiChiara:2017cjq, Sala:2017ihs, Ghosh:2017ber, Alok:2017sui, Alok:2017jaf, Wang:2017mrd, Bonilla:2017lsq, Bardhan:2017xcc, Crivellin:2017zlb, Fajfer:2015ycq, Bauer:2015knc, Barbieri:2015yvd, Das:2016vkr, Boucenna:2016qad, Becirevic:2016yqi, Bhattacharya:2016mcc, Chen:2017hir, Kumar:2018kmr, Blanke:2018sro, Crivellin:2018yvo, Crivellin:2019szf, Li:2019xmi, Li:2018rax, Popov:2019tyc}.  Remarkably, there appears to be a rather simple explanation for both the CC and NC anomalies in terms of a single vector leptoquark $U$ with SM quantum numbers $(3, 1, \frac{2}{3})$ that couples dominantly to left-handed quarks and leptons. A clear guide to the combined explanation of the anomalies may be found in Ref.~\cite{Buttazzo:2017ixm}. For a mass $m_U \sim 1~\tev$ and ${\cal O}(1)$ couplings to the third generation, the $U$ leptoquark can explain the $R(D^{(*)})$ and $R(K^{(*)})$ anomalies, at least for the central $q^2$ data.  Weak-scale states do not fully resolve the low-$q^2$ discrepancy, since a larger effect is required to modify the larger SM widths near the photon pole, but the $U$ leptoquark does also reduce the discrepancy for the low-$q^2$ data to roughly $1.7 \, \sigma$~\cite{Alok:2017sui}. 

The $U$ leptoquark does not, however, resolve the $(g-2)_{\mu}$ anomaly; it contributes at one loop, but this contribution is too small.  We must therefore introduce additional particles if we are also to explain the $(g-2)_{\mu}$ discrepancy.  Explanations in terms of additional weak-scale states, such as sleptons and gauginos~\cite{Feng:2001tr}, remain viable, but the implications of these explanations for experiments are very well known.  Alternatively, the $(g-2)_{\mu}$ anomaly could be resolved by light and very weakly coupled particles.  Dark photons with mass $\sim 10~\mev - 1~\gev$ were previously proposed as possible solutions~\cite{Boehm:2003hm,Pospelov:2008zw}, but these solutions are now excluded~\cite{Battaglieri:2017aum}.  However, other light-particle solutions remain viable.  For example, a light leptophilic scalar can contribute significantly to $(g-2)_{\mu}$ for large $\tan\beta \sim 200$, while its relatively weak hadronic couplings allow it to avoid stringent bounds~\cite{Batell:2016ove}.

In this work, we consider a different and novel light, weakly coupled particle solution to the $(g-2)_{\mu}$ problem: a light scalar $S$ with mass $m_S \sim 10 - 200~\mev$ that is an extension of the standard Type II 2HDM model.  The scalar $S$, which we will often refer to as the dark Higgs boson, couples to both leptons and quarks, but with couplings that are suppressed both by Yukawa couplings and by a small mixing parameter $\sin \theta$.  At the one-loop level, its contribution to $(g-2)_{\mu}$ is too small to resolve the anomaly. However, motivated by the leptoquark solution to the $B$ anomalies, we note that leptoquarks (as well as other TeV-scale particles) will generically induce an $S \gamma \gamma$ coupling, and this can resolve the $(g-2)_{\mu}$ anomaly through a two-loop Barr-Zee diagram.  In this way, the solutions to the $(g-2)_{\mu}$ and $B$ anomalies proposed here are connected.  (As an aside, we note that, for values of $m_S$ just below $2 m_{\mu}$, our explanation can also completely remove the discrepancy in the low-$q^2$ $R_{K^*}$ measurement, following a possibility noted previously in Ref.~\cite{Altmannshofer:2017bsz}.)

In addition to resolving longstanding anomalies, the proposed explanation predicts new signals.  In particular, given the light state $S$ and its couplings to electrons and photons, the model predicts new meson decays, such as $B \to K S$ and $K \to \pi S$, followed by $S \to e^+ e^-, \gamma \gamma$, leading to dilepton and diphoton signals that could be discovered in current and near-future experiments. The model also predicts exotic Higgs boson decays $h \to SS \to \gamma \gamma \gamma \gamma$, which may appear in detectors as a contribution to the $h \to \gamma \gamma$ signal.

In \secref{model}, we present the model, including the new fields we introduce and the relevant model parameters.  In \secref{muon}, we determine the parameter values that resolve the $(g-2)_{\mu}$ anomaly.  In \secref{B}, we then discuss constraints on the model from hadronic physics and show that a resolution to the $(g-2)_{\mu}$  and $B$ constraints exists in a viable region of parameter space.  The interesting implications for exotic $B$, $K$, and Higgs boson decays are discussed in \secref{newsignals}.  We summarize our conclusions in \secref{conclusions}.  \appsref{2hdm}{di-photon} contain details of the 2HDM model and the effective $S \gamma \gamma$ coupling, respectively.

%%%%%%%%%%%%%%%%%%%%%%%%%%%%%%%%%%%
\section{The Model}
\label{sec:model}
%%%%%%%%%%%%%%%%%%%%%%%%%%%%%%%%%%%

Our model is an extension of the Type II 2HDM.  The Type II 2HDM contains two Higgs doublets $H_u$ and $H_d$, which get vacuum expectation values (VEVs) $v_u$ and $v_d$ and give mass to the up-type and down-type fermions, respectively.\footnote{Although we will not be considering supersymmetry or supersymmetric states in this work, we note that the Type II 2HDM is the Higgs sector of the minimal supersymmetric standard model.}  We extend this by adding a singlet scalar $\phi$, which couples to the Higgs doublets through the portal interactions
\begin{equation}
V_{\text{portal}}= A \, (H_u^\dagger H_d + H_d^\dagger H_u )\phi  + \left[\lambda_u H_u^\dagger H_u + \lambda_d H_d^\dagger H_d  +  \lambda_{ud} (H_u^\dagger H_d + H_d^\dagger H_u) \right ]\phi \phi\ ,
\label{portal}
\end{equation}
where $CP$ conservation is assumed.
In this extension, we consider parameters such that $H_u$ and $H_d$ get VEVs, but $\phi$ does not.  After electroweak symmetry breaking, then, the trilinear scalar couplings mix the new scalar with the Higgs bosons of the 2HDM, and the quartic scalar couplings contribute to new Higgs boson decays $h \to \phi \phi$ and to the mass of the $\phi$.

More precisely, to determine the physical states of the theory, we minimize the full Higgs potential and diagonalize the mass matrices; for details, see \appref{2hdm}.  In the end, the physical states include the SM-like Higgs boson $h$ and the heavy Higgs bosons $H$, $A$, and $H^{\pm}$ of the 2HDM, but also a new real scalar, the dark Higgs boson $S$, with Lagrangian
\begin{equation}
\mathcal{L}_{S} = \frac{1}{2}(\partial_\mu S)^2 \! - \! \frac{1}{2}m_{S}^2 S^2 \! - \sin\theta  \tan \beta \! \sum_{f=d,l}  \! \frac{m_f}{v} \bar{f}f  S - \sin\theta' {\cot \beta} \! \sum_{f=u} \! \frac{ m_f }{v}\bar{f}f S  - \frac{1}{4} \kappa S F_{\mu \nu} F^{\mu \nu} ,
\label{L_S}
\end{equation}  
where $v \simeq 246~\gev$ and $\tan \beta = v_u / v_d$.  The couplings to fermions are inherited from the mixing of the dark Higgs boson with the 2HDM Higgs bosons: they are suppressed by Yukawa couplings, and the down-type couplings are enhanced by $\tan\beta$, while the up-type couplings are suppressed by $\cot\beta$. In addition, they are modified by the mixing angles $\sin \theta$ and $\sin\theta'$.  For weak portal interactions $A \ll m_h$ and large $\tan \beta$, these mixing angles can be written in terms of the physical Higgs boson masses.  As shown in \appref{2hdm}, the results are  
\bea
\sin\theta \approx -\frac{vA}{m_H^2}\ , \quad \sin\theta' \approx -\frac{2vA}{m_h^2}\left(1-\frac{m_h^2}{2m_H^2} \right) .
\label{mixing_angles}
\eea
The last term of \eqref{L_S} is an $S \gamma \gamma$ coupling governed by the parameter $\kappa$, which has dimensions of inverse mass.  This coupling is generically induced by heavy states, such as leptoquarks, as will be discussed in \secref{muon}. 

Finally, as discussed in \secref{introduction}, we add a vector leptoquark $U$ with SM quantum numbers $(3, 1, \frac{2}{3})$ and Lagrangian
\begin{equation}
\mathcal{L}_{U} = 
- \frac{1}{4} F_{\mu \nu}^{U} F^{U \mu \nu}  -  m^2_{U} U_{\mu} U^{\mu}  
- \left[ h_{ij}^{U} \left( \bar{Q}_{iL}\gamma^\mu L_{jL} \right)U_{\mu}  + \text{H.c.} \right] 
- g m_U S U_{\mu} U^{\mu} \, .
\label{L_U}
\end{equation}
The $U$ leptoquark's couplings to left-handed quarks and leptons resolve the $B$-meson anomalies. The leptoquark's couplings to right-handed quarks and leptons are constrained to be small~\cite{Datta:2019zca}. We have also included the leptoquark's couplings to $S$.  This interaction is allowed by all symmetries, but it will not play an important role in any of the phenomenology discussed below. As we will discuss later, we consider the $U$ leptoquark coupling to photons to be the same as the one between the $W$ boson and photons. Since the leptoquark is colored, it couples to gluons also \cite{Aebischer:2019mlg}. This coupling leads to their pair production at high energies, but it does not affect our phenomenology here.

In summary, the model we consider consists of a 2HDM model extended to include a light dark Higgs boson $S$ and a leptoquark $U$.  The leptoquark's couplings $h_{ij}^U$ are chosen to resolve the $B$ anomalies~\cite{Bhattacharya:2016mcc}. In addition to these, the parameters of the theory that are most relevant for the phenomenology we discuss below are
\begin{equation}
m_S, \tan \beta, \sin \theta, m_H, \kappa \ ,
\end{equation}
where $\tan \beta$, $\sin \theta$, and $m_H$ fully determine $\sin \theta'$ and the $S$ couplings to fermions, and $\kappa$ determines the $S$ couplings to photons.  We will be primarily interested in the parameter ranges $m_S \sim 10 - 200~\mev$, moderate to large $\tan\beta \sim 10-60$, small mixing angles $\sin \theta \sim 0.005$, $m_H \sim 1~\tev$, and $\kappa \sim (1~\tev)^{-1}$.

%%%%%%%%%%%%%%%%%%%%%%%%%%%%%%%%%%%
\section{Resolving the Muon Magnetic Moment Anomaly}
\label{sec:muon}
%%%%%%%%%%%%%%%%%%%%%%%%%%%%%%%%%%%

Given a 2HDM extended to include a dark Higgs boson $S$ and a vector leptoquark $U$ through the Lagrangian terms of \eqsref{L_S}{L_U}, respectively, we can now calculate the beyond-the-SM contributions to $(g-2)_{\mu}$.

\subsection{Dark Higgs boson contribution from effective \boldmath{$S \gamma \gamma$} coupling}

\begin{figure}[tbp]
\begin{center}
\includegraphics[width=0.35\textwidth]{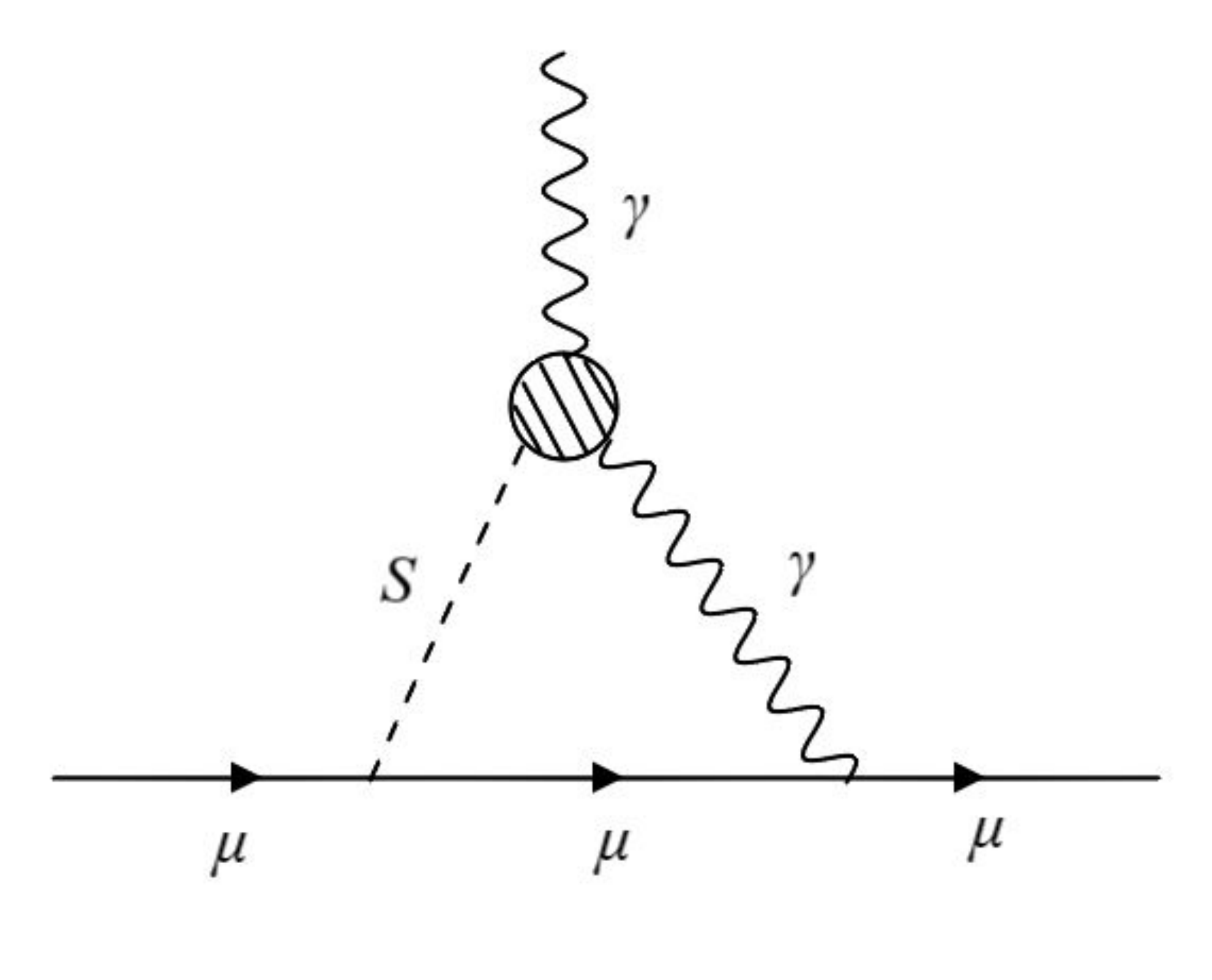}
\caption{Contribution of the effective $S\gamma\gamma$ coupling to $(g-2)_{\mu}$. 
\label{fig:Barr_Zee}
}
\end{center}
\end{figure}

\begin{figure}[tbp]
\begin{center}
\includegraphics[width=0.65\textwidth]{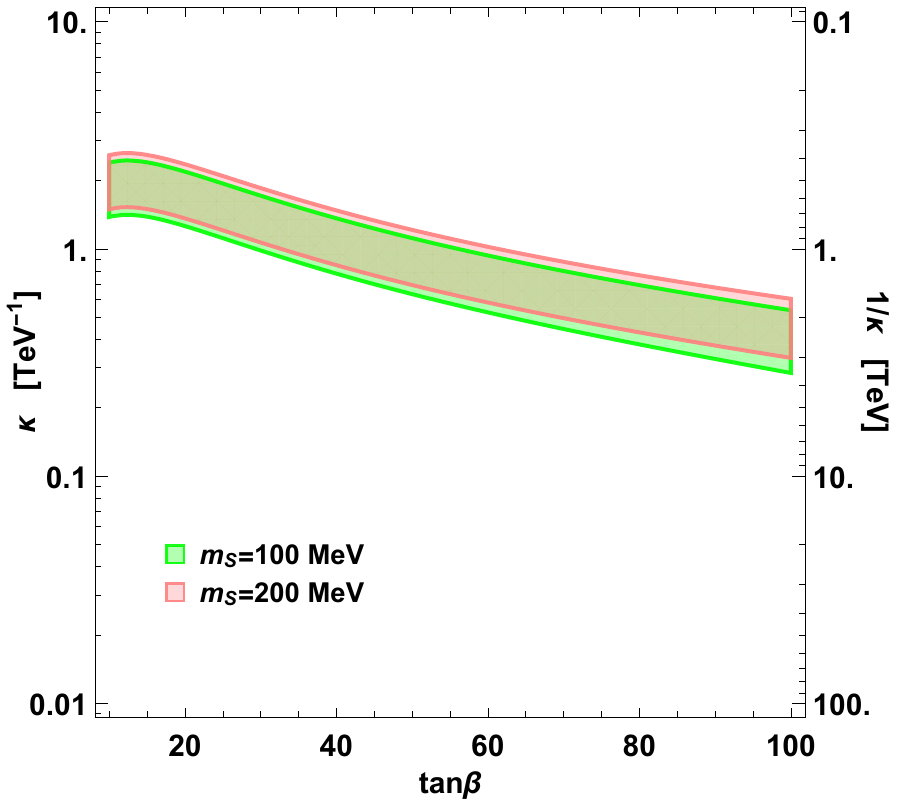}
\caption{The region of the $(\tan \beta, \kappa)$ plane where an effective $S\gamma \gamma$ coupling induces a Barr-Zee contribution to $(g-2)_{\mu}$ that enhances the theoretical prediction to be within $1 \, \sigma$ of the measured value.  The subdominant one-loop contribution from a virtual $S$ has also been included.  We fix $\sin \theta = 0.005$, $\Lambda = 2~\tev$, and show results for $m_S = 100~\mev$ and 200 MeV, as indicated. 
\label{fig:kappaparams}
}
\end{center}
\end{figure}

Let us first consider the dark Higgs boson contribution from the $S \gamma \gamma$ effective coupling shown in \figref{Barr_Zee}.  This contribution is dominated by the log-enhanced term~\cite{Davoudiasl:2018fbb} 
\begin{equation}
\Delta (g-2)_{\mu}^{S\gamma\gamma} \approx \frac{1}{4 \pi^2} \sin \theta \tan \beta \frac{m_{\mu}^2}{v} \kappa 
\ln \left( \frac{\Lambda}{m_S} \right) \ , 
\label{g-2}
\end{equation}
where $\Lambda$ is the cutoff scale, which we may take to be of the order of the mass of the particles that induce the effective $S \gamma \gamma$ coupling.   Parameters required to resolve the $(g-2)_{\mu}$ anomaly are presented in \figref{kappaparams}.   For dark Higgs mixing angle $\sin \theta \sim 0.005$ and $\tan\beta \sim 10 - 60$, we see that the effective coupling required is $\kappa \sim (1~\tev)^{-1}$.  In our calculations we also include the contribution to the  lepton anomalous magnetic moment at the one-loop level, which has been calculated to be~\cite{Chen:2015vqy}
\begin{equation}
\delta a_{\ell}^{(\text{1-loop)}}=\frac{g_\ell^2}{8\pi^2}\int_0^1 dz\frac{(1+z)(1-z)^2}{(1-z)^2+r^{-2} z} \ ,
\end{equation}
where $r=m_\ell/m_S$ and, in our case, $g_{\ell}=\sin\theta \tan\beta (m_\ell/ v)$.

\subsection{Dark Higgs boson contribution from \boldmath{$S \gamma \gamma$} coupling induced by $V$ leptoquarks}
\label{sec:Vcontributions}

How could such values of $\kappa$ be induced?  As an example, motivated by the effectiveness of leptoquarks for explaining the $B$ anomalies, we consider adding $N_{\text{LQ}}$ vector leptoquarks $V_i$, $i = 1, \ldots, N_{\text{LQ}}$, with Lagrangians
\begin{equation}
\mathcal{L}_{V_i} = 
- \frac{1}{4} F_{\mu \nu}^{V_i} F^{V_i \mu \nu}  -  m^2_{V_i} V_{i \mu} V^{\mu}_i 
- \left[ h_{jk}^V \left( \bar{Q}_{jR}\gamma^\mu L_{kR} \right) V_{i \mu}  + \text{H.c.} \right] 
- g_{V_i} m_{V_i} S V_{i\mu} V^{\mu}_i \ ,
\label{L_V}
\end{equation}
where for simplicity we add only leptoquarks with SM quantum numbers $(3, 1, \frac{5}{3})$ and assume that their couplings to right-handed quarks and leptons are  identical.

Assuming small couplings $h_{jk}^V$, the leading way in which these $V_i$ leptoquarks contribute to $(g-2)_{\mu}$ is by inducing an $S \gamma \gamma$ coupling, which then contributes through a Barr-Zee diagram.  The Barr-Zee contribution to $(g-2)_{\mu}$ with a $W$ boson in the loop has been calculated in Ref.~\cite{Ilisie:2015tra} in the context of 2HDMs. As leptoquarks are not gauge bosons, there might be ambiguities in the leptoquark two-loop contribution. For an $O(1)$ estimate of this contribution, we model the effect of this leptoquark loop by the $W$ loop. We find that the leptoquark contributions to $(g-2)_{\mu}$ are always positive---that is, in the right direction---and they induce an effective $S \gamma \gamma$ coupling parameter
\begin{equation}
\kappa = \frac{\alpha_{\text{EM}}}{4\pi} \sum_{i=1}^{N_{\text{LQ}}} \frac{N^c Q^2 g_{V_i}}{m_{V_i}}  F_W(4m_{V_i}^2/m_S^2) \, ,
\label{effectivekappa}
\end{equation}   
where $\alpha_{\text{EM}} \simeq 1/137$; $N^c = 3$ and $ Q = \frac{5}{3}$ are the number of colors and electric charge of the leptoquarks $V_i$, respectively; $g_{V_i}$ parametrizes the $S V_i V_i$ coupling in \eqref{L_U}; and $F_W$ is a loop function defined in Ref.~\cite{Marciano:2011gm}. 

\begin{figure}
\begin{center}
\includegraphics[width=0.45\textwidth]{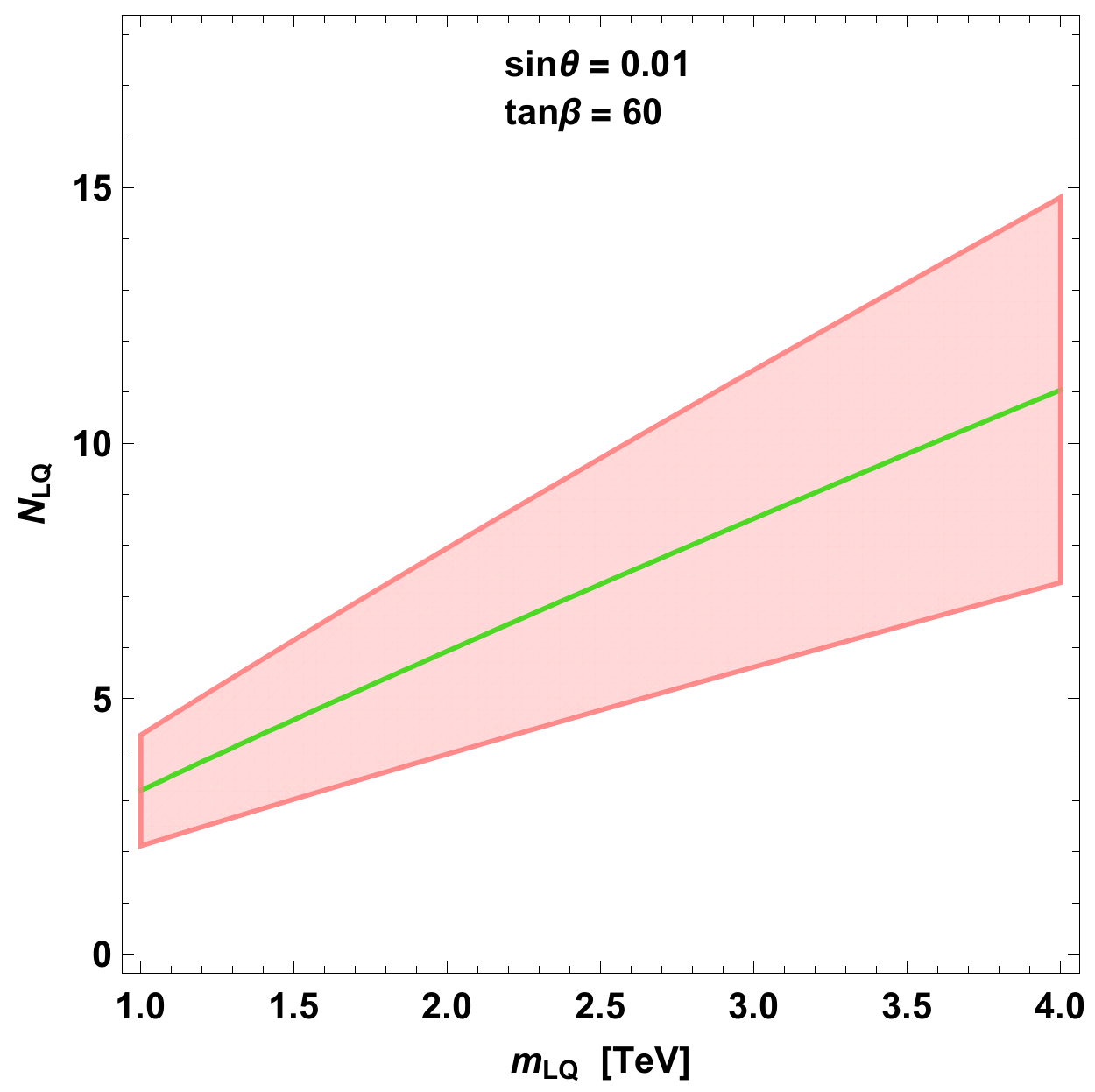} \quad
\includegraphics[width=0.45\textwidth]{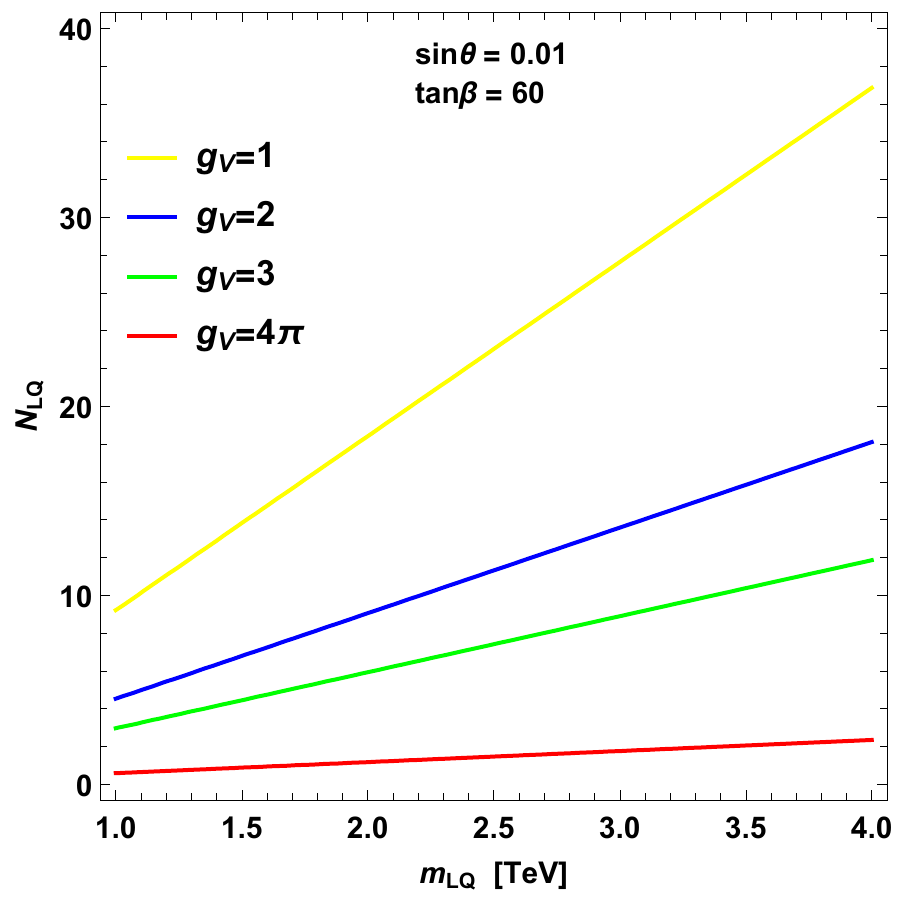} \\[10pt]
\includegraphics[width=0.45\textwidth]{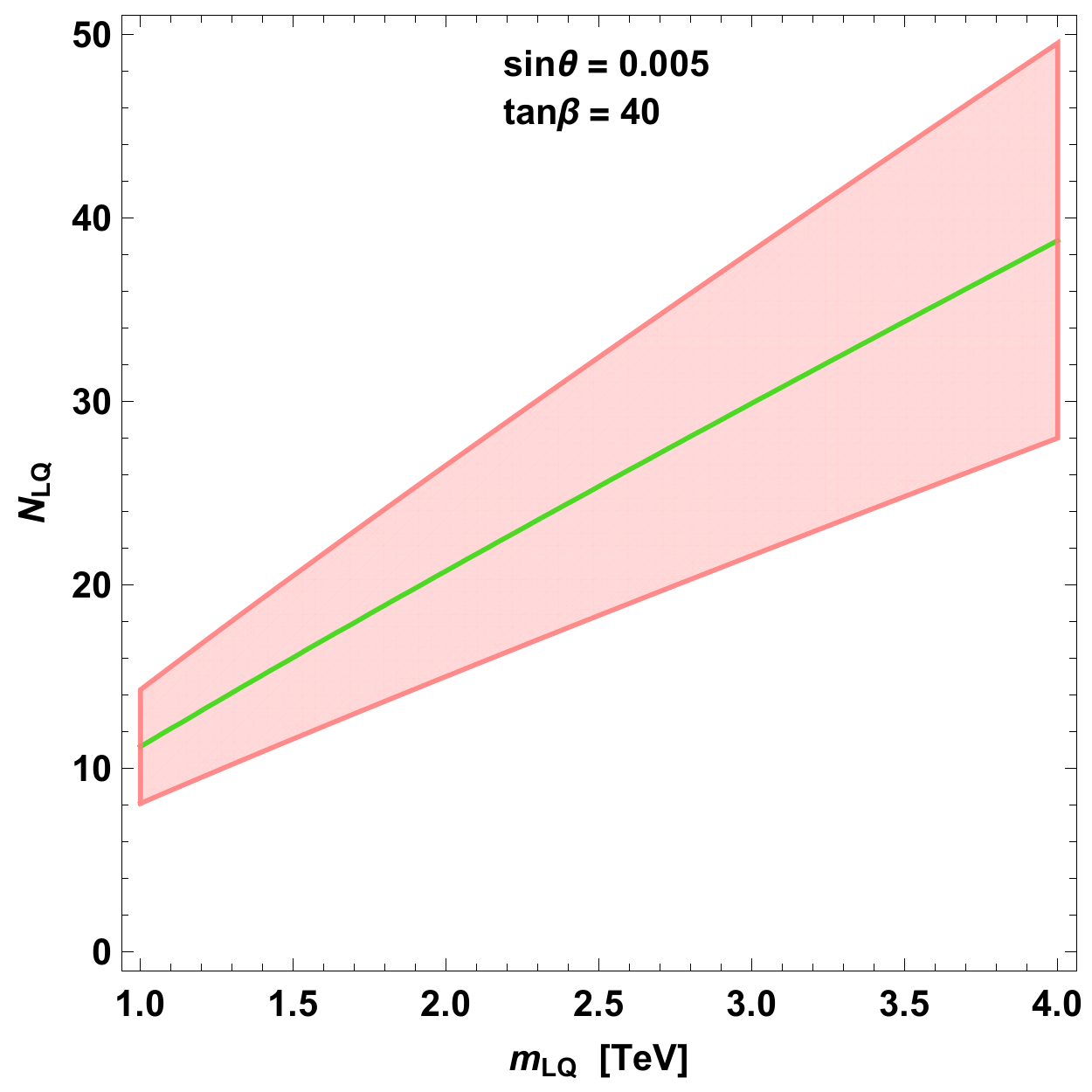} \quad
\includegraphics[width=0.45\textwidth]{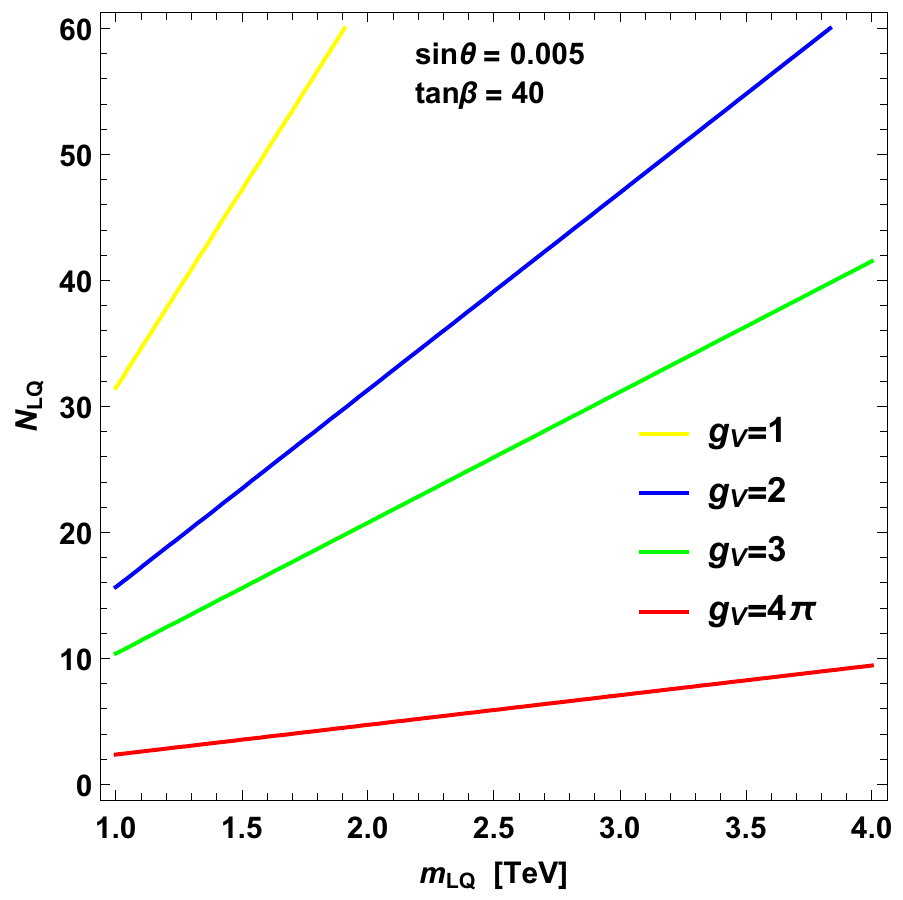}
\caption{The region of the $(m_{\text{LQ}}, N_{\text{LQ}})$ plane where $N_{\text{LQ}}$ vector leptoquarks $V_i$ with mass $m_{\text{LQ}}$ and SM quantum numbers $(3, 1, \frac{5}{3})$ induce an effective $S\gamma \gamma$ coupling that resolves the $(g-2)_{\mu}$ anomaly.  In all panels, we set $m_S = 100~\mev$.  In the upper and lower panels, we fix $(\sin \theta, \tan\beta) = (0.01, 60)$ and $(0.005, 40)$, respectively.  For the left panels, we set $g_V =3$ and show the bands where the $(g-2)_{\mu}$ discrepancy is reduced to $1 \, \sigma$.  For the right panels, we consider the several values of $g_V$ indicated and plot the lines on which the theoretical prediction for $(g-2)_{\mu}$ exactly matches its experimentally measured value.  [In the upper- and lower-right panels, the induced couplings are $\kappa \simeq (3.2~\tev)^{-1}$ and $(0.9~\tev)^{-1}$, respectively.]
\label{fig:leptoquarkparams}
}
\end{center}
\end{figure}

For large leptoquark masses $m_{V_i} \gg m_S$, the loop function is $F_W \simeq 7$. In the simple case where we have $N_{\text{LQ}}$ copies of degenerate leptoquarks with mass $m_{V_i} = m_{\text{LQ}}$ and coupling $g_{V_i} = g_V$, \eqref{effectivekappa} reduces to
\begin{equation}
\kappa \simeq 0.034 \frac{N_{\text{LQ}} \, g_V}{m_{\text{LQ}}} \ .
\end{equation}
Setting $g_V = 3$ and requiring $\kappa \approx \tev^{-1}$, the mass and number of leptoquarks required to resolve the $(g-2)_{\mu}$ anomaly are related by $m_{\text{LQ}} \approx N_{\text{LQ}} \, (100~\gev)$.  The required parameters are shown graphically in \figref{leptoquarkparams}.

We see that it is not difficult to induce an effective $S \gamma \gamma$ coupling large enough to resolve the $(g-2)_{\mu}$ anomaly.  For the $\tan\beta = 60$ case shown, with even just $N_{\text{LQ}} = 5$ leptoquarks with mass $m_{\text{LQ}} = 2~\tev$, which is currently viable, one can reduce the discrepancy in $(g-2)_{\mu}$ to $1 \, \sigma$.  Alternatively, one can achieve the same result with $N_{\text{LQ}} = 10$ leptoquarks with mass $m_{\text{LQ}} = 4~\tev$, which is likely challenging even for searches at the High Luminosity LHC. For the $\tan \beta = 40$ case shown, one requires roughly twice as many leptoquarks, but the number is still not very large.  Of course, the assumed new physics that is necessarily light is the dark Higgs boson $S$.  This will have interesting observable consequences, as we discuss in \secref{B}.

\subsection{\boldmath{$U$} Leptoquark contribution}

\begin{figure}[tbp]
\begin{center}
\includegraphics[width=0.3\textwidth]{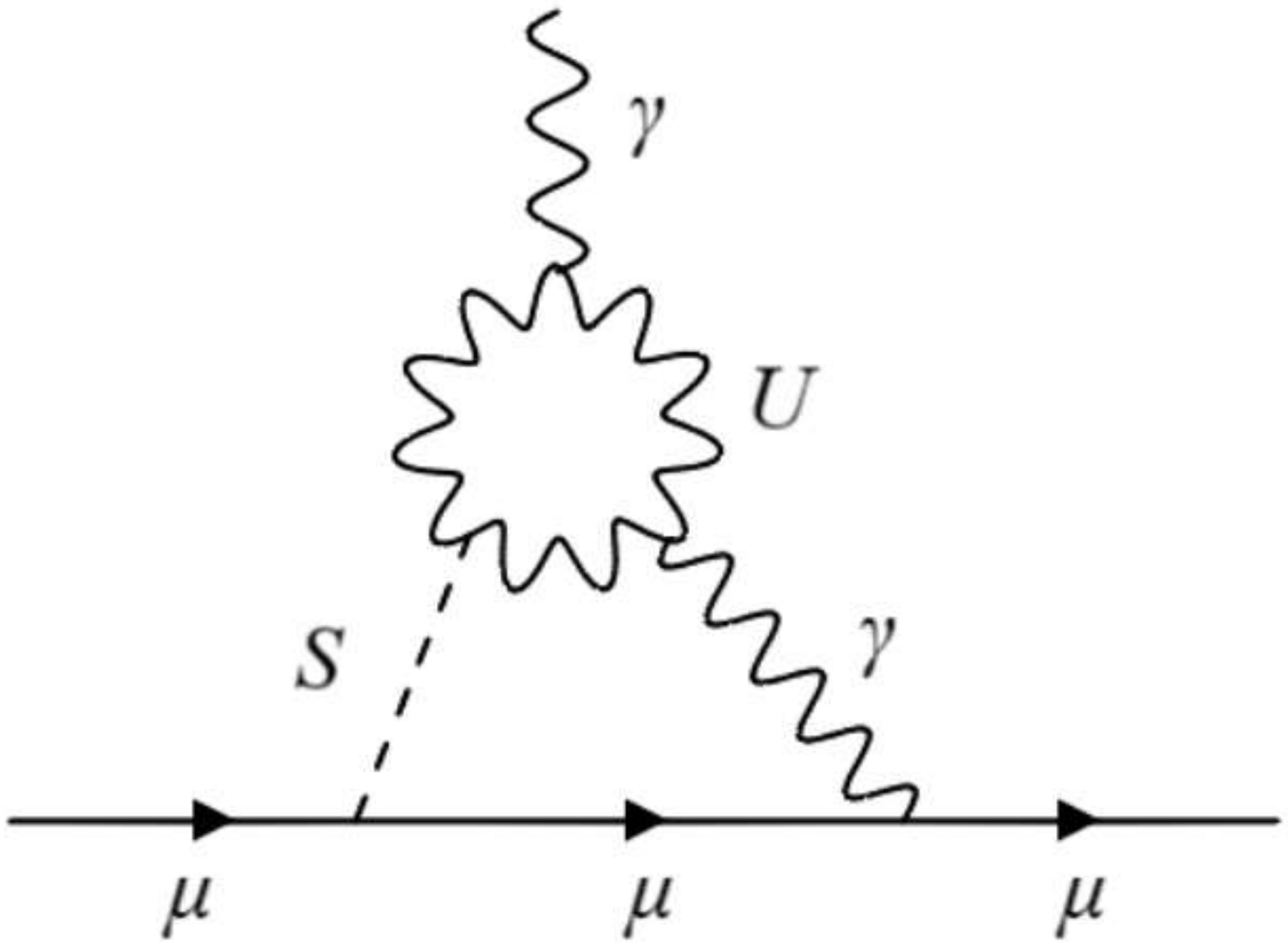} \quad
\includegraphics[width=0.3\textwidth]{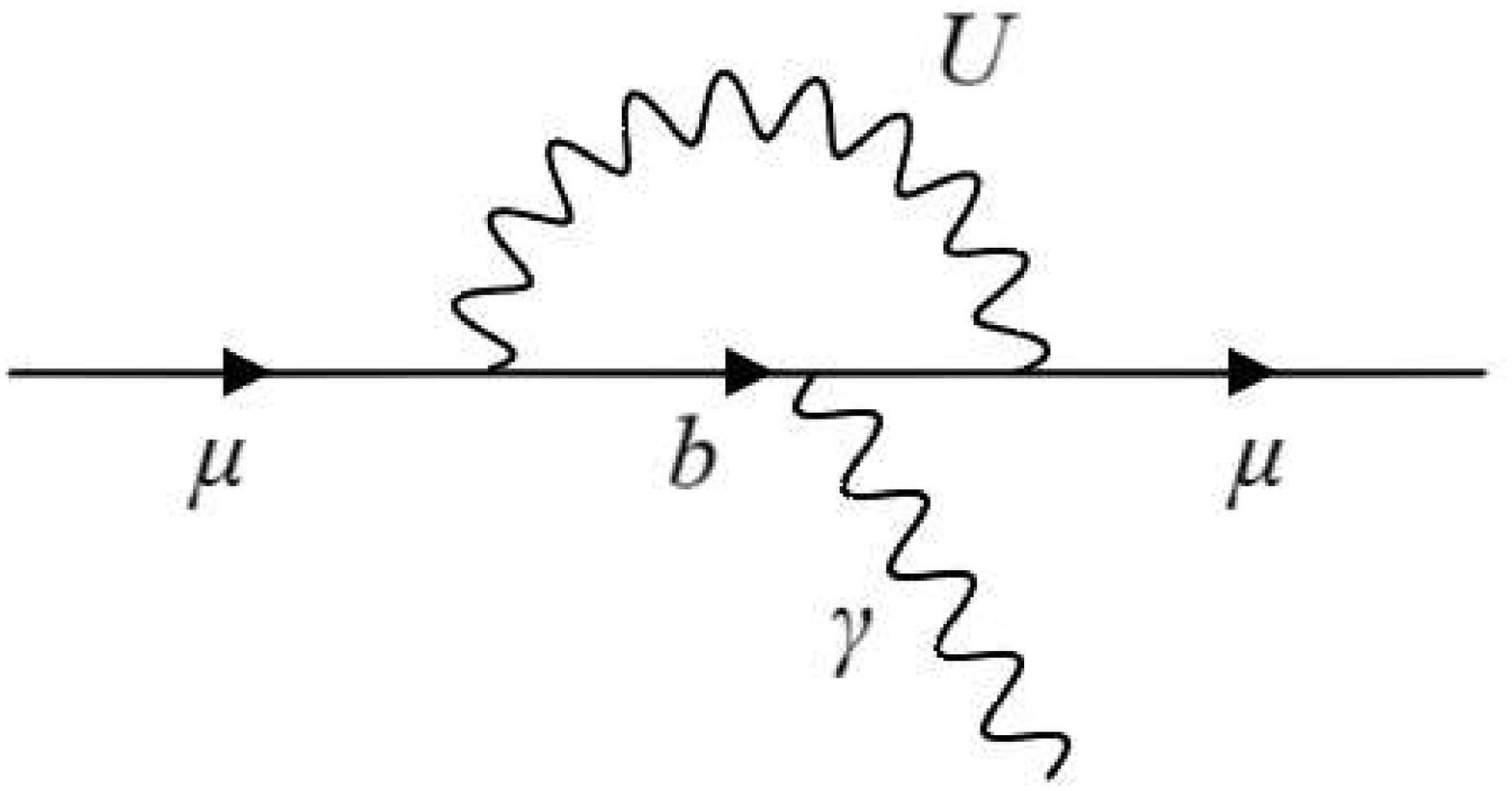} \quad
\includegraphics[width=0.3\textwidth]{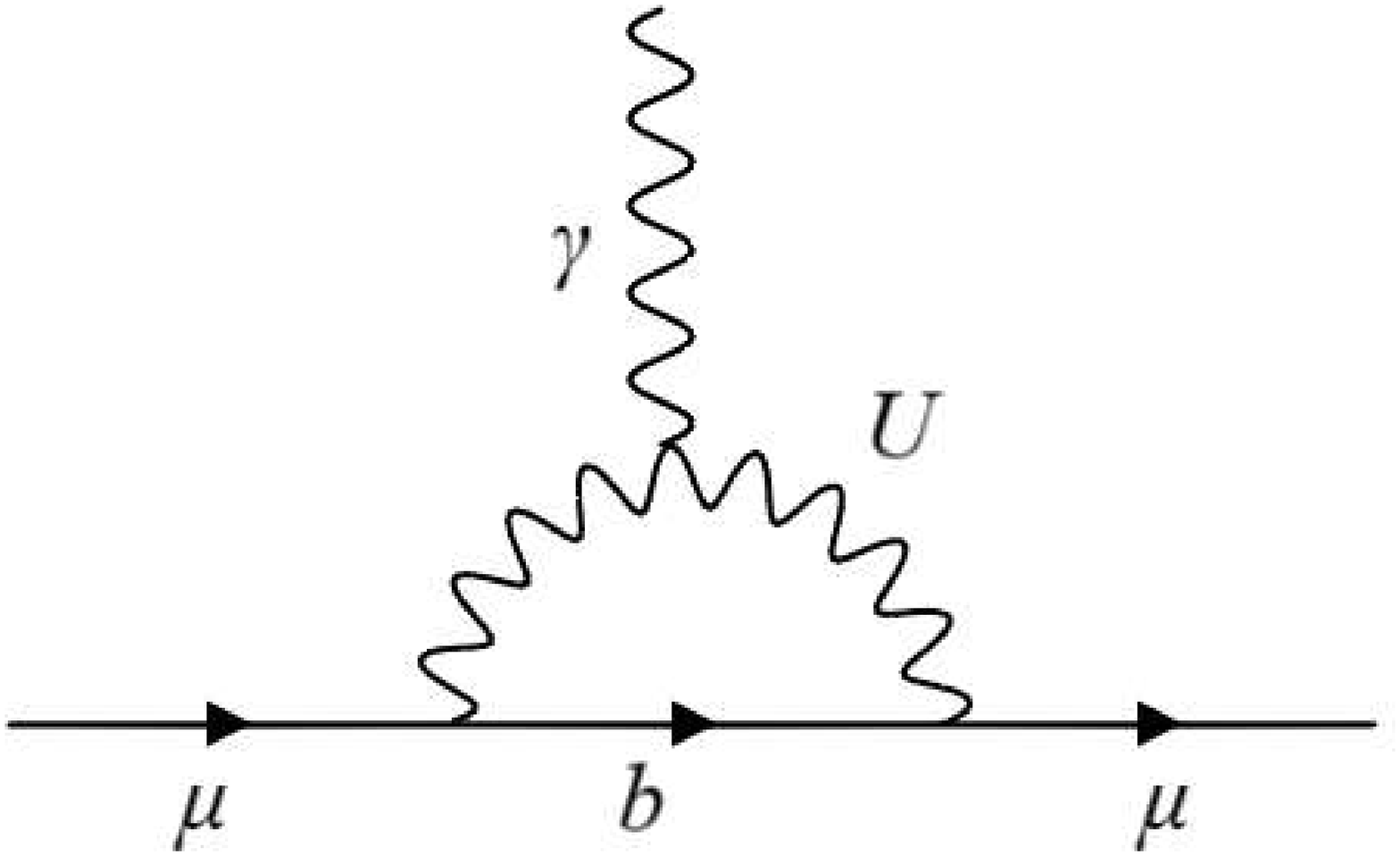}
\caption{$U$ leptoquark contributions to $(g-2)_{\mu}$.  Left: Two-loop Barr-Zee diagram involving also the dark Higgs boson $S$.  Center and right: One-loop diagrams that are independent of the dark Higgs boson.
\label{fig:U1_g-2}
}
\end{center}
\end{figure}

In addition to the contributions to $(g-2)_{\mu}$ mediated by the dark Higgs boson and independent of the $U$ leptoquark, there are also the contributions that depend on the $U$ leptoquark shown in \figref{U1_g-2}.  These include the two-loop Barr-Zee contribution from a $S \gamma \gamma$ coupling mediated by the $U$ leptoquark, similar to those discussed above for $V$ leptoquarks in \secref{Vcontributions}, and also two one-loop contributions independent of the dark Higgs boson.

The two-loop Barr-Zee diagram's contribution is as discussed above.  The contribution of a single $U$ leptoquark with mass $\sim \tev$ is not sufficient to raise the theoretical prediction for $(g-2)_{\mu}$ to the experimental value.  

In addition, however, there are the one-loop contributions from the coupling of $U$ to the muon and down-type quarks, $h_{i \mu }^{U} \bar{d}_{iL} \gamma^\nu \mu_L U_\nu$, where $i = d, s, b$. These contributions to $(g-2)_{\mu}$ are~\cite{Queiroz:2014zfa}
\begin{equation}
\Delta (g-2)_{\mu}^{U}  = \sum_{i = d, s, b}  -\frac{N^c (h_{i \mu }^{U})^2}{16\pi^2} \left( \frac{4m_\mu^2}{3m_{U}^2}Q_i - \frac{5m_\mu^2}{3m_{U}^2}Q_{U} \right),
\end{equation} 
where $N^c=3$ is the number of colors, and $Q_{i} = -\frac{1}{3}$ and $Q_U = -\frac{2}{3}$ are the electric charges of the down-type quarks and the $U$ leptoquark. Substituting these charges and the value for the muon mass, we find
\begin{equation}
\Delta (g-2)_{\mu}^{U}  = \sum_{i=d,s,b} -1.4 \times 10^{-10} (h_{i \mu }^{U})^2 \left( \frac{\tev}{m_U} \right) ^2 \ .
\label{g-2_LQ}
\end{equation}
This contribution is of the wrong sign to explain the $(g-2)_{\mu}$ anomaly and depends on the couplings $h_{i \mu}^{U}$. In particular, the couplings $h_{b \mu }^{U}$ and  $h_{s \mu }^{U}$ contribute to $b \to s \mu^+ \mu^-$ and are used to explain the $R(K^*)$ and $\bsmumu$ anomalies~\cite{Bhattacharya:2016mcc, Kumar:2018kmr}. As we show in the next section, however, the couplings $h_{i \mu }$ have small enough values that we can ignore the one-loop contribution to $(g-2)_\mu$. In summary, then, the $U$ leptoquark contributions to $(g-2)_{\mu}$ are negligible in our model and do not modify our discussion about the $V$ leptoquark requirements to resolve the $(g-2)_{\mu}$ anomaly.

%%%%%%%%%%%%%%%%%%%%%%%%%%%%%%%%%%%
\section{Resolving the \boldmath{$B$} Anomalies and Hadronic Constraints}
\label{sec:B}
%%%%%%%%%%%%%%%%%%%%%%%%%%%%%%%%%%%

\subsection{The \boldmath{$U$} leptoquark and \boldmath{$B$} anomalies}

The couplings of the $U$ leptoquark in \eqref{L_U} can resolve all the $B$ anomalies. Let us start with the $\bsmumu$ anomalies, which include the $\RK$ and $\RKstar$ measurements. The procedure to fit for new physics  is the following.  The $\bsmumu$ transitions are defined via an effective Hamiltonian with vector and axial vector operators:
\bea
H_{\rm eff} &=& - \frac{\alpha G_F}{\s \pi} V_{tb} V_{ts}^*
      \sum_{a = 9,10} ( C_a O_a + C'_a O'_a ) \ , \nn\\
O_{9(10)} &=& [ {\bar s} \gamma_\mu P_L b ] [ {\bar\mu} \gamma^\mu (\gamma_5) \mu ] \ ,
\label{Heff}
\eea
where the $V_{ij}$ are elements of the Cabibbo-Kobayashi-Maskawa (CKM) matrix, and the primed operators are obtained by replacing $L$ with $R$. The Wilson coefficients include both SM and new physics contributions: $C_a = C_{a,\, \SM} + C_{a,\, \NP}$. One now fits to the data to extract  $C_{a,\NP}$.  There are several scenarios that give a good fit to the data, and the results of recent fits can be found in Refs.~\cite{Alok:2019ufo, Ciuchini:2019usw, Aebischer:2019mlg, Alguero:2019pjc,Datta:2019zca, Kowalska:2019ley}. One of the popular solutions is  $C_{9,\, \NP}^{\mu\mu} = - C_{10,\, \NP}^{\mu\mu}$, which can arise from the  tree-level exchange of the $U$ leptoquark in \eqref{L_U}.  Following the results of Ref.~\cite{Datta:2019zca}, fitting to the $\bsmumu$ data constrains the central values of the $U$ couplings to satisfy
\bea
h_{b \mu }^{U} \, h_{s \mu }^{U} &=& 8 \times 10^{-4} \ . 
\label{Ucons}
\eea
The framework to explain all the $B$ anomalies,  including both the CC and the NC anomalies, involves the $U$ leptoquark coupling to the third-generation quarks and leptons in the gauge basis with $O(1)$ coupling, $ h_{b \tau }^{U} \sim 1$~\cite{Bhattacharya:2016mcc}. As one moves from the gauge to the mass basis, for the quarks and leptons, the couplings $h_{b \mu }^{U}$ and  $h_{s \mu }^{U}$ are generated. Hence, one has the hierarchy $h_{b \tau }^{U} \sim 1 >h_{b \mu }^{U}> h_{s \mu }^{U}>h_{d \mu }^{U}$. Using the allowed values of $h_{b \mu }^{U} \sim 0.1 - 0.6$~\cite{Bhattacharya:2016mcc} and \eqref{Ucons}, we see that the one-loop $U$ contribution to $(g-2)_{\mu}^{U}$ in \eqref{g-2_LQ} cannot resolve the $(g-2)_{\mu}$ discrepancy.  The $(g-2)_{\mu}$ anomaly therefore requires additional new physics, such as the $S$ boson discussed in \secref{muon}.

\subsection{Hadronic constraints}

In this model, the $S$ boson inherits its couplings from the Higgs boson, and so it necessarily couples to both leptons and hadrons.  The lepton couplings, specifically the muon coupling, are desired to resolve the $(g-2)_{\mu}$ anomaly.  Here we begin to examine the implications of the hadronic couplings, which may either constrain the model or lead to predictions of interesting new signals.

Particularly stringent are constraints on FCNC processes, since couplings like $bsS$ are induced through a penguin loop. Integrating out the $W$-top loop induces the effective $bsS$ vertex~\cite{Batell:2009jf}
\begin{equation}
\mathcal{L}_{bs}= \frac{\sin\theta'}{v \tan\beta} \frac{3\sqrt{2}G_F m_t^2 V^*_{ts}V_{tb}}{16 \pi^2}m_b \bar{s}P_R b S + \text{H.c.} \ ,
\end{equation}
where the factor $\frac{\sin\theta'}{v \tan\beta}$ comes from the top quark coupling to $S$.  By the same loop process, but replacing $b$ and $s$ quarks with $s$ and $d$ quarks, respectively, the $sdS$ vertex is also generated.  Note that the FCNC amplitude depends on the mixing angle $\sin {\theta'}$ in Eq.~(\ref{mixing_angles}), which is suppressed by $m_h^2$, while the $(g-2)_\mu$ in Eq.~(\ref{g-2}) is controlled by the mixing angle $\sin \theta$ in Eq.~(\ref{mixing_angles}), which is suppressed by  $m_H^2$. If a higher value  of $m_H$ is compensated by a larger value of the mixing parameter $A$ to keep the same $\sin \theta$, then $\sin \theta'$ can become too large and be inconsistent with FCNC data.

\begin{figure}
\begin{center}
\includegraphics[width=0.65\textwidth]{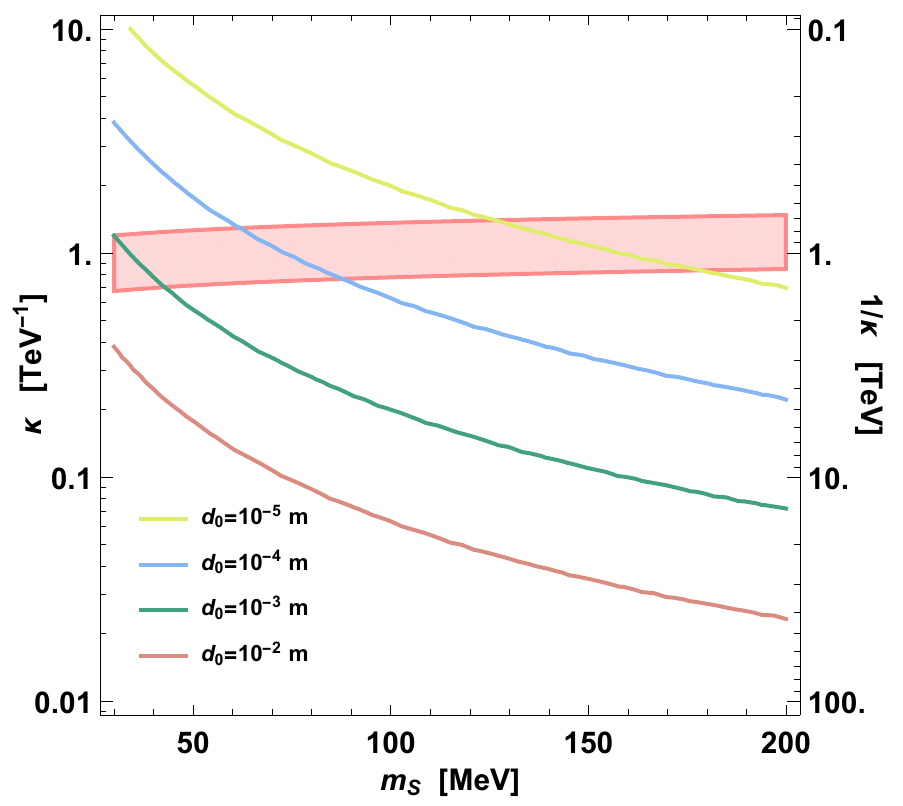}
\caption{Contours of constant flight distance (excluding the boost factor) ($d_0 = c \tau_0$) of the light scalar $S$ in the $(m_S, \kappa)$ plane. We fix $\sin\theta = 0.005$ and $\tan\beta = 40$.  In the pink shaded region, the $(g-2)_{\mu}$ anomaly is reduced to $1 \, \sigma$. }
\label{fig:lifetime}
\end{center}
\end{figure}

The FCNC interactions will induce two-body decays $B \to K^{(*)}S$  and $K \to \pi S$. To determine the signature of these processes, it is important to determine how the $S$ decays. For $m_S \sim 10 - 200~\mev$, the possible decays are $S \to e^+e^-, \gamma \gamma$. In \figsref{lifetime}{BtoKseecons}, we show the $S$ lifetime and branching fraction to $e^+e^-$, respectively.  We see that for most of the parameters of interest, the $S$ flight distance (excluding the boost factor) is $c\tau_0 \alt 1~\text{mm}$, and so the $S$ decay is effectively prompt.  We also see that the dominant decay is to diphotons, with $BR(S \to e^+ e^-) \sim 10^{-5} - 10^{-3}$ in the parameter region of interest.

\begin{figure}
\begin{center}
\includegraphics[width=0.65\textwidth]{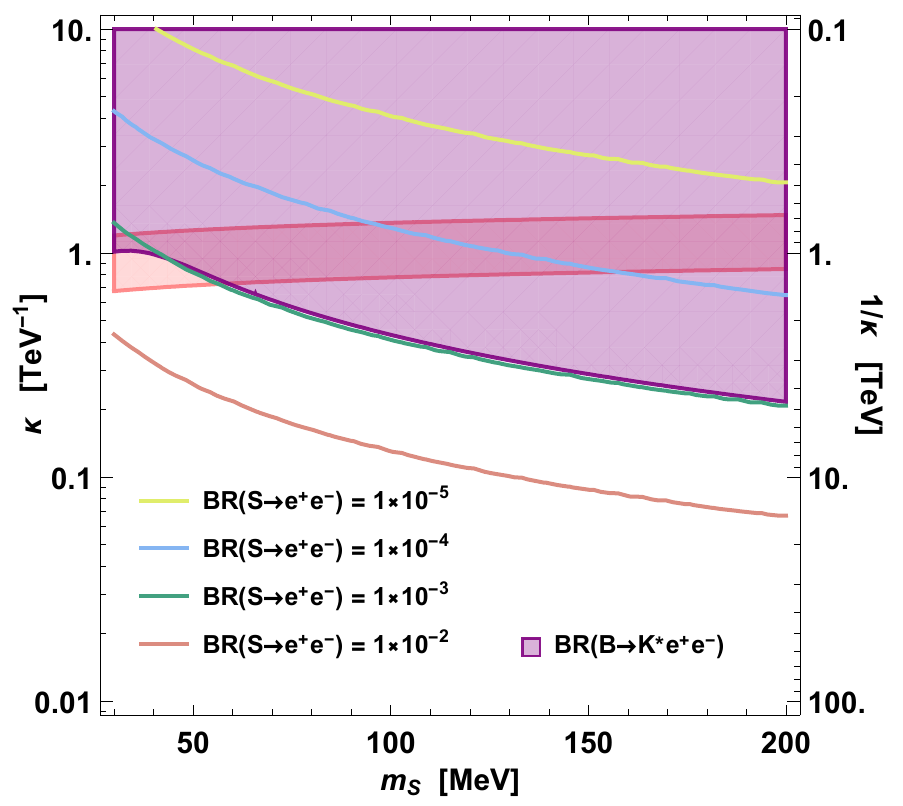}
\caption{Contours of constant branching fraction $BR(S \to e^+ e^-)$ in the $(m_S, \kappa)$ plane. We fix $\sin\theta = 0.005$ and $\tan\beta = 40$.  In the pink shaded region, the $(g-2)_{\mu}$ anomaly is reduced to $1 \, \sigma$, and in the purple shaded region, $BR(B \to  K^{*}  e^+ e^-)$ is within $1 \, \sigma$ of its measured value.}
\label{fig:BtoKseecons}
\end{center}
\end{figure}

We now determine the rates for the two-body decays $B \to K^{(*)}S$  and $K \to \pi S$. For the two-body decays $B \to K^{(*)}S$ we have~\cite{Datta:2017pfz, Datta:2017ezo}
\begin{equation}
\label{BtoKS}
BR(B \to K S)=\frac{g_{bs}^2 f_0^2(m_S^2) (m_B^2-m_K^2)^2 |\vec{p}_K| \tau_B}{32 \pi m_B^2 (m_b-m_s)^2}
\end{equation}
and 
\begin{equation}
\label{BtoKsS}
BR(B \to K^* S)=\frac{g_{bs}^2 A_0^2(m_S^2) |\vec{p}_{K^*}|^3 \tau_B}{8\pi (m_b + m_s)^2} \ ,
\end{equation}
where $m_b$ and $m_s$ are the bottom and strange quark masses, respectively; $f_0$ and $A_0$ are form factors, which are taken from Refs.~\cite{Straub:2015ica,Ball:2004ye}; and $g_{bs}$ is the flavor-changing $b \to s$ coupling with the normalization $\mathcal{L}_{bs}= g_{bs} \bar{s}P_R b S$. Given the prompt $S$ decays to $e^+e^-$ and $\gamma \gamma$,  we have $ BR(B \to  K^{(*)} e^+ e^-)$  dominantly coming from $BR(B \to  K^{(*)} S) BR(S \to e^+ e^- ) $ and $ BR(B \to  K^{(*)} \gamma \gamma)$  dominated by $BR( B \to  K^{(*)} S) BR(S \to \gamma \gamma) $. One can extend this to $K$ decays also. 

We now discuss constraints from $B$ and $K$ decays on this model.  In this subsection, we will consider a variety of nonleading constraints and show that they are far from excluding the favored parameter space of this model.  These observables are listed in Table~\ref{Tab: other_obs} and are the following:

\begin{table}[]
\begin{tabular}{| l | c | c |}
\hline
Observable                  & \begin{tabular}[c]{@{}l@{}}New scalar contribution \\ $\sin\theta=0.005$, $\tan\beta=40$\end{tabular} & Existing constraints/measurements  \\ 
\hline
$BR(B \to KS)$                       & $1.7 \times 10^{-4}$       &  $ < 10\% $                                                                                                               \\ \hline
$BR(B \to K^* S)$                    & $1.7 \times 10^{-4}$                                                                                                                                                                        
&   $ < 10\% $
 \\ \hline
$BR(B_s \to \mu^+\mu^- )$            &  $4.2 \times 10^{-14}$                                          
&     $(3.0 \pm 0.4) \times 10^{-9}$                     
\\ \hline
$BR(B_s \to \gamma \gamma)$              & $7.4 \times 10^{-11}$
& $ < 3.1 \times 10^{-6}$                                                                   
\\ \hline
$\Delta M^{NS}_{B_s}$                & $-2.5 \times 10^{-17}~\gev$
&  $ < 1.7 \times 10^{-12}~\gev$                                                                   \\ \hline
$\Delta M^{NS}_{K}$                  & $-6.3 \times 10^{-24}~\gev$
&    $ < 5.9 \times 10^{-18}~\gev$                                                                   \\ \hline
$BR(K^+ \to \mu^+ \nu e^+ e^-)$      & $3.3 \times 10^{-14}$
&  $ (7.81 \pm 0.23) \times 10^{-8} $                                                                    \\ \hline
$BR(K^{\pm} \to \pi^{\pm} e^+ e^-)$     & $8.7 \times 10^{-11}$
&   $ (3.11 \pm 0.12) \times 10^{-7} $                                                                    \\ \hline
$BR(K_S \to \gamma \gamma)$              & $3.3 \times 10^{-16} $
& $  (2.63 \pm 0.17) \times 10^{-6}$                                                                   
\\ \hline
$BR(K_L \to \gamma \gamma)$              & $3.2 \times 10^{-14} $
& $ (5.47 \pm 0.04) \times 10^{-4}$                                                                   
\\ \hline
$\delta (g-2)_e$						&	$6.3 \times 10^{-14}$        					&  $(-87 \pm 36) \times 10^{-14}$
\\ \hline
\end{tabular}
\caption{Values of the contribution of the new scalar $S$ to various meson observables. We fix the dark scalar mass to $m_S=100~\mev$. References for the experimental constraints are given in the text.}
\label{Tab: other_obs}
\end{table}

\begin{itemize}

\item $B$ total decay width: In the first two rows of Table~\ref{Tab: other_obs}, we require that $BR(B \to K^{(*)}S)$ not exceed the uncertainty in the SM prediction of the width of the $B$ meson, which we take to be around 10\%~\cite{Lenz:2014jha}.

\item $B_s$ decay: The process $B_s \to \mu^+\mu^-$ is mediated by an $s$-channel dark Higgs boson $S$, where the matrix element is  $\mathcal{M}_{B_s \to \mu^+\mu^-}=\frac{g_{bs}g_{\mu}}{m_{B_s}^2-m_S^2} (\bar{s}P_Rb) \, (\bar{\mu}\mu)$.
We use {\tt flavio}~\cite{Straub:2018kue} to calculate the contribution of the light scalar $S$ to this decay mode. 
The branching ratio of this decay is measured to be $(3.0 \pm 0.4)\times 10^{-9}$~\cite{Tanabashi:2018oca}.  The process $B_s \to \gamma \gamma$ is also mediated by an $s$-channel $S$. The SM prediction for $BR(B_s \to \gamma \gamma)$ is around $5 \times 10^{-7}$~\cite{Reina:1997my}, and there exists an experimental upper bound of $3.1 \times 10^{-6}$~\cite{Tanabashi:2018oca} for this observable. The branching ratio of the decay in terms of the effective $S \gamma \gamma$ coupling $\kappa$ is
\begin{equation}
BR(B_s \to \gamma \gamma)=\frac{ |g_{bs}|^2 |\kappa|^2}{64 \pi}\frac{f_{B_s}^2 m_{B_s}^7}{m_b^2 (m_{B_s}^2-m_S^2)^2}  \tau_{B_s} \ .
\end{equation}

\item $B_s$ and $K$ mixing: In the SM, the $B_s$ mass difference is $\Delta M_{B_s}^{\text{SM}}=(17.4 \pm 2.6)~\text{ps}^{-1}$~\cite{Bhattacharya:2016mcc}. We require that the new scalar contribution not exceed the SM uncertainty. The expression for the mass difference due to the new scalar is~\cite{Atwood:1996vj, Datta:2017ezo}
\begin{equation}
\Delta M_{B_s}^{\text{NS}}=-\frac{5}{24}\frac{g_{bs}^2}{m_{B_s}^2-m_S^2}f_{B}^2 m_{B_s} \ .
\end{equation}
 We use a similar equation for the $K-\bar{K}$ mixing mass difference and use the experimental value $\Delta M_K^{\text{exp}}=(52.93 \pm 0.09)\times 10^{8}~\text{s}^{-1}$~\cite{Tanabashi:2018oca}. 

\item $K$ decay: The  rare decay $K^+ \to \mu^+ \nu e^+e^-$ has been measured by the NA48/2 Collaboration to be $BR(K^+ \to \mu^+ \nu e^+e^-) = (7.81 \pm 0.23)\times10^{-8}$~\cite{KhoriauliNA48/2}, where the measurement is restricted to the kinematic region with $m_{e^+e^-}\geq 140~\text{MeV}$. To study this decay mode, we calculate the branching ratio of the decay $K \to \mu \nu_\mu S$, where the scalar particle $S$ is radiated off the muon leg~\cite{Carlson:2012pc}.  The total branching ratio is then determined through
\begin{equation}
BR(K^+ \to \mu^+ \nu_\mu e^+ e^-)=BR(K^+ \to \mu^+ \nu_\mu S)BR(S \to e^+ e^-) \ .
\end{equation}

The $K^{\pm} \to \pi^{\pm} e^+ e^-$ mode also has been measured by the NA48/2 Collaboration to be $BR(K^{\pm} \to \pi^{\pm} e^+ e^-)=(3.11 \pm 0.12)\times 10^{-7}$~\cite{Batley:2009aa}.  For this process we find the two-body decay rate $K^{\pm} \to \pi^{\pm} S$, and the branching ratio of the desired process is determined by
\begin{equation}
BR(K^{\pm} \to \pi^{\pm} e^+ e^-)=BR(K^{\pm} \to \pi^{\pm} S)BR(S \to e^+ e^-) \ .
\end{equation} 

\item $K_{S,L}$ decays:  The decays $K_{S,L} \to \gamma\gamma$ are mediated through $s$-channel dark Higgs bosons $S$, just as in the case $B_s \to \gamma\gamma$ discussed above. The new contributions to these decay modes and their Particle Data Group values~\cite{Tanabashi:2018oca} are presented in Table~\ref{Tab: other_obs}.

\item Last, although not a hadronic constraint, we also list the model prediction for $(g-2)_e$. Just as there is a Barr-Zee contribution to $(g-2)_{\mu}$, there is an analogous Barr-Zee contribution to $(g-2)_e$. In contrast to the muon case, the measured value for $(g-2)_e$ is smaller than the SM prediction, and so our model's contribution to $(g-2)_e$ is in the wrong direction. However, as can be seen in Table~\ref{Tab: other_obs}, the contribution to $(g-2)_e$ is very small, and does not significantly worsen the agreement between theory and experiment.  

\end{itemize}

We see that none of the constraints listed in Table~\ref{Tab: other_obs} is a significant constraint on the model.  In the next section, we will consider the leading constraints, which do constrain parts of the model parameter space, but also provide interesting predictions for signals that could be seen in the near future.

%%%%%%%%%%%%%%%%%%%%%%%%%%%%%%%%%%%
\section{New Signals of the Model}
\label{sec:newsignals}
%%%%%%%%%%%%%%%%%%%%%%%%%%%%%%%%%%%

\subsection{\boldmath{$B \to  K^{(*)} e^+ e^-$}}

As noted above, the model contributes to the decay $B \to  K^{(*)}  e^+ e^-$ with branching fraction $BR(B \to  K^{(*)}  e^+ e^-)  = BR(B \to  K^{(*)} S) BR(S \to e^+ e^- ) $. The region of the $(m_S, \kappa)$ parameter space that is consistent with the measured value of $BR(B \to  K^{(*)}  e^+ e^-) = (3.1^{+0.9}_{-0.8} \, {}^{+0.2}_{-0.3} \, \pm 0.2) \times 10^{-7}$~\cite{Aaij:2013hha} is shown in Fig.~\ref{fig:BtoKseecons}, along with the region in which the $(g-2)_{\mu}$ anomaly is resolved.  We see that the existing constraint on $BR(B \to  K^{(*)}  e^+ e^-)$ excludes the very lowest values of $m_S \sim 10~\mev$, but most of the parameter space is allowed.  Future measurements of $BR(B \to  K^{(*)}  e^+ e^-)$ with increased sensitivity may therefore see a deviation predicted by this model.  There is also a measurement of the inclusive $ B \to X_s e^+e^-$ decay~\cite{Lees:2013nxa} for $ 0.1 < m_{e^+e^-}^2<2.0~\gev^2 $, but this is outside the $m_S$ range we consider and so cannot be used to constrain our model.

\subsection{\boldmath{$B \to  K^{(*)} \gamma\gamma$}}

\begin{figure}
\begin{center}
\includegraphics[width=0.65\textwidth]{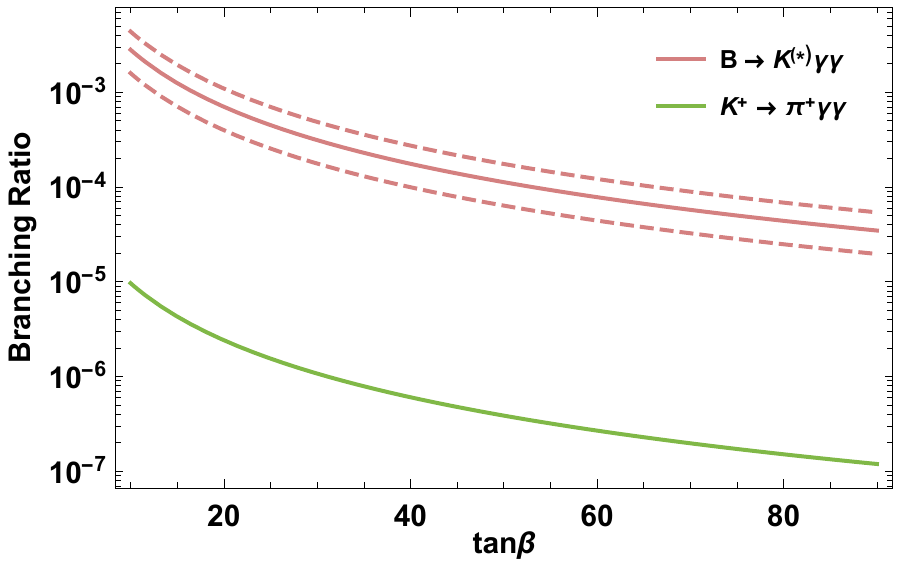}
\caption{The values of the branching fractions for the decays $B \to K^{(*)}\gamma\gamma$ and $K^+ \to \pi^+\gamma\gamma$. The branching fractions for $B \to K\gamma\gamma$ and $B \to K^*\gamma\gamma$ are essentially identical.  The dashed bands correspond to the $2 \, \sigma$ variations of the $B \to K^{(*)}$ form factors. We fix $\sin\theta = 0.005$ and $m_S = 100~\mev$. }
\label{fig:BtoK2gamma}
\end{center}
\end{figure}

As the $S$ decays almost always to diphotons, another important signal for the $S$ state is from $B \to K^{(*)}\gamma\gamma$ decays.  In \figref{BtoK2gamma}, we show the predictions for $B \to  K^{(*)} \gamma\gamma$. The predictions depend on the $B \to  K^{(*)}$ form factors $f_0$ and $A_0$ mentioned above. We show the range of the predictions as we vary the form factors within $2 \, \sigma$ of the quoted uncertainty. It should be noted that the form factors are not from first-principle QCD calculations, and so one should keep that in mind when discussing uncertainties in the form factors. The predictions for  $ B \to  K \gamma\gamma  $ and $ B \to  K^{*} \gamma\gamma$ are almost identical, and they range from roughly $1 \times 10^{-4}$ to $ 3 \times 10^{-4}$ for $\tan \beta=40$.  

Because the $\gamma \gamma$ comes from a light $S$, for a sufficiently low $m_S$, the two $\gamma$ may be collinear and look like a single $\gamma$. One of the $\gamma$ may also be soft, in which case again the $ 2 \gamma$ will look like a single $\gamma$. Hence, experimentally one should check the $ B \to K^{(*)} \gamma$ signal carefully to look for signs of a diphoton resonance. We should also point out that our predictions for the $ B \to  K^{*} \gamma\gamma $ rates should be considered as ballpark estimates, as one can choose a more general 2HDM model to relax the branching ratio predictions. If the mass of the  $S$ is close to the $\pi^0$ mass, the final states for { $ B \to  K^{(*)} \pi^0 $ and  { $ B \to  K^{(*)} S  $, with both $\pi^0$ and $S$ decaying to $\gamma \gamma$, are the same, and one will have to consider carefully adding the two contributions. As nonleptonic decays are very difficult to calculate it will be difficult to detect the presence of the $S$ particle in this case or obtain constraints on the model from the $ B \to  K^{(*)} \pi^0 $ measurement.  In the SM, the nonresonant decay $ B \to X_s \gamma \gamma$ has a branching ratio around $ 4 \times 10^{-7}$~\cite{Reina:1997my}, where the photons are required to have an energy greater than 100 MeV. Also, in Ref. \cite{Hiller:2004wc}, a study of the short-distance effects in $B \to K^{(*)} \gamma \gamma$ decays, together with the resonant contributions, is presented.  At present, the observed $ B \to K^{(*)} \gamma \gamma$ signals come only from known resonances, but analyses of the currently unexplored nonresonant regions could yield signals of the dark Higgs boson $S$.

\subsection{\boldmath{$ K \to  \pi \gamma\gamma$} }
 
In \figref{BtoK2gamma}, we also show the predicted branching ratios for $K^+ \to  \pi^+ \gamma\gamma$. For $\tan\beta = 40$, the prediction is approximately $6 \times 10^{-7}$.  If the $S$ mass is near the $\pi^0$ mass, the $K^+ \to  \pi^+ \gamma\gamma$ decay will be swamped by the $K^+ \to \pi^+ \pi^0$ decay, which has a branching ratio of about 21\%~\cite{Tanabashi:2018oca}. Away from the $\pi^0$ resonance, there is a measurement of the nonresonant $K^+ \to  \pi^+ \gamma\gamma$ decay with  branching ratio $(1.01 \pm 0.06) \times 10^{-6}$~\cite{Tanabashi:2018oca}, but this measurement is obtained by combining measurements made for diphoton invariant masses above the range of $S$ masses we consider.  The predictions of this model could be tested by future measurements with this sensitivity, but for diphoton masses between 10 and 200 MeV. 

For the neutral kaons, the model predictions for $\sin\theta=0.005$, $\tan\beta=40$, and $m_S=100~\mev$ are $BR(K_L \to  \pi^0 S) = 4 \times 10^{-7}$ and $BR(K_S \to  \pi^0 S) = 4 \times 10^{-9}$.  The much smaller branching ratio for $K_S$ is largely due to the $K_S$ having a much shorter lifetime than $K_L$, while the $K^+$ and $K_L$ lifetimes are of the same order.  The measured branching ratios are $BR(K_L \to  \pi^0 \gamma\gamma)= (1.273 \pm 0.033) \times 10^{-6}$ and $BR(K_S \to  \pi^0 \gamma\gamma)= (4.9 \pm 1.8) \times 10^{-8}$~\cite{Tanabashi:2018oca}. Again, the model predictions are not far from current sensitivities and predict a sharp signal with diphoton mass equal to $m_S$.

\subsection{ \boldmath{$\lowercase{h} \to \gamma \gamma \gamma \gamma$} and implications for \boldmath{$\lowercase{h} \to \gamma \gamma$} }
\label{sec:higgs}

The model discussed here may also modify Higgs boson decays through the process $h \to SS$, followed by $S \to \gamma \gamma$.\footnote{The model also predicts heavy Higgs boson decays $H \to SS$, but the branching ratio for this is very small, of the order of $10^{-6}$.} Since the SM Higgs boson is much heavier than the scalar $S$, the two photons from $S$ decay are boosted and highly collimated.  Therefore, the decay $h \to S(\to\gamma\gamma)S(\to \gamma\gamma)$ contributes to the $h \to \gamma \gamma$ signal~\cite{Dobrescu:2000jt}. We can calculate the couplings appearing in the $\frac{1}{2} g_{hSS} h SS $ interaction in terms of the parameters of the potential and mixing parameters. The resulting branching ratio is
\begin{equation}
BR(h \to SS) = \frac{g_{h SS}^2}{32 \pi m_{h} \Gamma_{h} } \sqrt { 1- \frac{4 m_S^2}{ m_{h}^2}  } \ .
\end{equation}
	
The signal strengths measured by CMS and ATLAS are $\mu^{\gamma\gamma}=1.18^{+0.17}_{-0.14}$~\cite{Sirunyan:2018ouh} and $\mu^{\gamma\gamma}=1.06^{+0.14}_{-0.12}$~\cite{ATLAS-CONF-2018-028}, respectively. By a naive combination of these two measurements, we find $\mu^{\gamma\gamma}=1.11 \pm 0.10$. (We averaged the CMS and ATLAS measurements to $\mu^{\gamma\gamma}=1.18 \pm 0.16$ and $\mu^{\gamma\gamma}=1.06 \pm 0.13$, respectively.)

In the parameter region of our interest in the model, we can find values for parameters of the potential such that the addition of the process $ h \to SS \to \gamma \gamma \gamma \gamma$ to the SM rate of $h \to \gamma \gamma$ does not exceed the measured signal strength. As an example, for $\sin\theta=0.005$ and $\tan\beta=40$, and taking $m_{du}=200~\gev$, $\lambda_1 = 0.6$, $\lambda_2 = 0.3$, $\lambda_{345} = 2.8$, $\lambda_d = -0.3$, $\lambda_u = 0.0005 $, and $\lambda_{ud} = 0.005$, the signal strength becomes $\mu^{\gamma\gamma} \approx 1.08$. Of course, this also implies that as the experimental constraints on $\mu^{\gamma\gamma}$ become more precise, a deviation from the SM expectation may appear.\footnote{As noted below \eqref{portal}, after electroweak symmetry breaking, the quartic interactions contribute to the $\phi$ mass.  For the quartic coupling values given here, we require some fine-tuning between this contribution and the bare mass for the mass of the physical scalar $S$ to be in the desired range $m_S \sim 10 - 200~\mev$.}

%%%%%%%%%%%%%%%%%%%%%%%%%%%%%%%%%%%
\section{Conclusions}
\label{sec:conclusions}
%%%%%%%%%%%%%%%%%%%%%%%%%%%%%%%%%%%

In this study, we have proposed a concrete model that resolves both the $(g-2)_{\mu}$ and $B$-meson anomalies, which are currently among the leading discrepancies between SM predictions and experimental data. The model is a Type II 2HDM model, such as the Higgs sector of the minimal supersymmetric model, extended to include a light dark Higgs boson $S$, a leptoquark $U$, and additional leptoquarks $V$.  The $U$ leptoquark resolves the $B$ anomalies, and the $V$ leptoquarks generate a $S \gamma \gamma$ coupling.  This coupling  induces a two-loop Barr-Zee contribution to $(g-2)_{\mu}$, which is shown in \figref{Barr_Zee}.  The model makes interesting predictions 
 for exotic signals that can be looked for in current and upcoming data. Our proposed resolution to the $(g-2)_{\mu}$ problem requires either a large number of LQs or a large coupling, or both, and  if there is a large coupling, it  could blow up just above the TeV scale, requiring a number of additional states in any UV-complete theory.  An UV completion of our model is beyond the scope of this work, but we believe that in any UV framework, the essential features of our model will remain valid.

For dark Higgs mass $m_S \sim 100~\mev$ and dark Higgs mixing angle $\sin \theta \sim 0.005$, $\tan\beta \sim 40$, and $N_{\text{LQ}} \sim 10$ $V$ leptoquarks with masses at the TeV scale, the correction resolves the $(g-2)_{\mu}$ anomaly.  The introduction of a new light scalar $S$ has many possible effects on SM meson phenomenology.  We have checked that all current bounds on $K$ and $B$ properties, as well as the current constraint on $(g-2)_e$, are respected for the parameters that solve the $(g-2)_{\mu}$ and $B$ meson anomalies; see Table~\ref{Tab: other_obs}.  

In the near future, however, there are measurements that could uncover beyond-the-SM effects and provide evidence for this model.  In particular, the dark Higgs boson is light enough to be produced in meson decays, and it then decays through $S \to e^+ e^-, \gamma \gamma$. The $S$ boson has $c\tau \sim 0.01 - 1~\text{mm}$, and so for most model parameters the decay is indistinguishable from prompt, which yields interesting new dielectron events from $B \to K^{(*)} e^+ e^-$ with $m_{e^+e^-} = m_S$ and diphoton signals from $B \to K^{(*)} \gamma \gamma$ and $K \to \pi \gamma \gamma$ with $m_{\gamma \gamma} = m_S$. The branching ratios for some of these modes are shown in \figsref{BtoKseecons}{BtoK2gamma}.  In all cases, the predicted branching ratios are not far from current sensitivities, although current measurements typically explore ranges of $m_{e^+e^-}$ and $m_{\gamma\gamma}$ outside the considered range of $m_S$.  As examples, the model predicts values $BR(B \to K^{(*)} \gamma \gamma) \sim 10^{-4}$ and $BR(K^+ \to \pi^+ \gamma \gamma), BR(K_L \to \pi^0 \gamma \gamma)\sim 10^{-6}$.  Provided the $S$ is not too degenerate with the neutral pion $\pi^0$, these signals could be observed above background in the near future---for example, at Belle II, providing a motivation to look for these exotic diphoton modes and an avenue for testing this model.  More generally, these decay modes test many models where the $(g-2)_{\mu}$ anomaly is resolved by a two-loop Barr-Zee contribution generated by a light $S$ with an $S \gamma \gamma$ coupling.

In addition, there are potentially observable contributions to exotic Higgs decays $h \to S S \to \gamma \gamma \gamma \gamma$, which, given that the $S$ is very light, typically lead to signals indistinguishable from $h \to \gamma \gamma$.  For the desired model parameters, the contribution to $h \to \gamma \gamma$ is within current constraints, but improved measurements could uncover a deviation from SM predictions.  Of course, electromagnetic calorimeters with extremely fine spatial resolution that could differentiate photons separated by opening angles of $\theta \sim m_S/m_h \sim \text{mrad}$ would be able to distinguish the $\gamma \gamma \gamma \gamma$ signal from the $\gamma \gamma$ signal, which would provide a smoking gun signal of new physics.

%%%%%%%%%%%%%%%%%%%%%%%%%%%%%%%%%%%
\acknowledgments
%%%%%%%%%%%%%%%%%%%%%%%%%%%%%%%%%%%

We thank W.~Altmannshofer, T.~Browder, L.~Cremaldi, R.~Harnik, and T. M. P.~Tait for discussions.  A. D. and S. K. are supported in part by NSF Grant No.~PHY-1414345. A. D. thanks the hospitality of the T. D. Lee Institute, where part of the work was completed. J. L. F. is supported in part by NSF Grants No.~PHY-1620638 and No.~PHY-1915005 and in part by Simons Investigator Award No. 376204. The work of J. K. is financially supported by NSERC of Canada.

\appendix

\section{Calculation of \boldmath{$S$} Couplings in Terms of 2HDM Model Parameters}
\label{sec:2hdm}

We now explicitly calculate the parameters in the Lagrangian in \eqref{L_S}, following the analysis of Ref.~\cite{Batell:2016ove}.  We start with the Type II 2HDM with the Yukawa couplings
\begin{equation}
-\mathcal{L}_Y = \bar{L}^0 Y^0_e H_d e^0_R + \bar{Q}^0 Y^0_d H_d d^0_R +\bar{Q}^0Y^0_u\tilde{H}_u U^0_R + \text{H.c.}
\end{equation}
Here the superscript denotes the quantities that are in flavor space.

We write the scalar potential as
\begin{equation}
V(H_d,H_u,\phi)=V_{\text{2HDM}}(H_d,H_u)+V_{\phi}(\phi)+V_{\text{portal}}(H_d,H_u,\phi) \ ,
\end{equation}
where 
\begin{eqnarray}
V_{\text{2HDM}} &=& m_{dd}^2 H_d^{\dagger}H_d + m_{uu}^2 H_u^{\dagger}H_u - m_{du}^2(H_d^{\dagger}H_u+H_u^{\dagger}H_d) + \frac{\lambda_1}{2}(H_d^\dagger H_d)^2+\frac{\lambda_2}{2}(H_u^\dagger H_u)^2 \nonumber \\
         && +\lambda_3(H_d^\dagger H_d)(H_u^\dagger H_u)+\lambda_4(H_d^\dagger H_u)(H_u^\dagger H_d)+\frac{\lambda_5}{2}\left[(H_d^\dagger H_u)^2+(H_u^\dagger H_d)^2 \right] \\
         V_\phi&=& B \phi + \frac{1}{2}m_0^2\phi^2+\frac{A_\phi}{2}\phi^3+\frac{\lambda_\phi}{4}\phi^4 \\
V_{\text{portal}} &=& A \, (H_u^\dagger H_d + H_d^\dagger H_u )\phi  + \left[\lambda_u H_u^\dagger H_u + \lambda_d H_d^\dagger H_d  +  \lambda_{ud} (H_u^\dagger H_d + H_d^\dagger H_u) \right ]\phi \phi\ .
\end{eqnarray}

After each doublet obtains a VEV, we write the neutral real components of the doublets as $H_i=v_i + \rho_i$, where $i=d, u$. After expanding the potential, the elements of the mass matrix of the $CP$-even scalars in the $(\rho_d,\rho_u,\phi)$ basis are
\begin{eqnarray}
M_{11}^2 &=& m_{du}^2\tan\beta + \lambda_1 v^2 \cos^2\beta \\
M_{22}^2 &=& m_{du}^2\cot\beta + \lambda_2 v^2 \sin^2\beta  \\
M_{12}^2 &=& -m_{du}^2+\lambda_{345}v^2\cos\beta \sin\beta \\
M_{13}^2 &=& v A \sin\beta  \\
M_{23}^2 &=& v A \cos\beta  \\
M_{33}^2 &=& m_0^2 + v^2 \lambda_d \cos\beta^2 + v^2 \lambda_u \sin\beta^2 + 2v^2 \lambda_{ud}\cos\beta \sin\beta   \ ,
\end{eqnarray}
where $\lambda_{345}=\lambda_3+\lambda_4+\lambda_5$, and $v_d$ and $v_u$ are the VEVs of the two doublets $H_d$ and $H_u$, with $\tan\beta=v_u/v_d$ and $v_d^2+v_u^2=v^2=(246~\gev)^2$. 

We assume $A \ll v,m_{du}$, so we can consider the portal terms as small perturbations. In this case, we diagonalize the mass matrix perturbatively, where the nonperturbed mass matrix is the usual 2HDM mass matrix. We define the mixing matrix that diagonalizes the mass matrix  as
\begin{equation}
\left(\begin{array}{c} \rho_d \\ \rho_u \\ \phi \end{array}\right) \approx 
\left({\begin{array}{ccc} -\sin\alpha & \cos\alpha & \delta_{13} \\
						   \cos\alpha & \sin\alpha & \delta_{23} \\
						   \delta_{31} & \delta_{32} & 1 \end{array}}\right) 
\left(\begin{array}{c} h \\ H \\ S \end{array}\right) \ ,
\end{equation}
where $\delta_{ij}$'s are small mixing angles that mix the light scalar with the other two scalars of the 2HDM. When we diagonalize the mass matrix of the 2HDM, the parameter $\alpha$ satisfies the usual equation 
\begin{equation}
\tan 2\alpha=\frac{2M_{12}^2}{M_{11}^2-M_{22}^2} \ ,
\end{equation}
and the masses of the two $CP$-even Higgs bosons are given by
\begin{equation}
m_{h,H}^2=\frac{1}{2}\left[M_{11}^2+M_{22}^2 \mp \sqrt{(M_{11}^2-M_{22}^2)^2+4(M_{12}^2)^2}  \right] .
\end{equation}
To determine expressions for the $\delta_{ij}$'s, we write the mass matrix as
\begin{equation}
M^2=\left({\begin{array}{ccc} M_{11}^2 & M_{12}^2 & 0     \\
						      M_{12}^2 & M_{22}^2 & 0      \\
						        0      &    0     & M_{33}^2  \end{array}}\right)
+   \left({\begin{array}{ccc}          0      &    0             & vA \sin\beta     \\
						               0      &    0             & vA \cos\beta      \\
						     vA \sin\beta & vA \cos\beta &   0     \end{array}}\right) \ ,
\end{equation} 
where the second matrix is considered as a small perturbation. Below, we use the shorthand notation $s_\beta=\sin\beta$ and $c_\beta=\cos\beta$. 

We require the lighter Higgs $h$ to have SM-like couplings to gauge bosons and fermions, so that we have $\beta - \alpha = \pi/2$. Assuming $M_{33} \ll m_h, m_H$, and writing $\alpha=\beta-\pi/2$, we find that the small mixing parameters are
\begin{eqnarray}
\label{deltas}
\delta_{13} &=& -\frac{2vA s^3_\beta}{m_h^2}\left[ \frac{m_h^2}{2m_H^2}+\cot^2\beta \left( 1-\frac{m_h^2}{2m_H^2} \right) \right] \nonumber  \\
\delta_{23} &=& -\frac{2vA}{m_h^2}s^2_\beta c_\beta 
\left[1-\frac{m_h^2}{2m_H^2}(1-\cot^2\beta)   \right] \nonumber \\
\delta_{31} &=& \frac{vA s_{2\beta}}{m_h^2}   \nonumber \\
\delta_{32} &=& -\frac{vA c_{2\beta}}{m_H^2} \ . 
\end{eqnarray}

In the Yukawa sector, after rotating to the mass basis and defining the mass matrices of fermions, the interaction terms between the physical light scalar $S$ and the fermions become
\begin{equation}
-\mathcal{L}_{ffS}=\left( \frac{\delta_{13}}{v c_\beta}\bar{e} M_e e  + \frac{\delta_{13}}{v c_\beta}\bar{d} M_d d + \frac{\delta_{23}}{v s_\beta} \bar{u} M_u u \right) S \, ,
\end{equation}
where the $M_f$'s are the diagonal mass matrices of the fermions.  To better compare with SM Higgs couplings, we write these couplings as
\begin{equation}
-\mathcal{L}_{ffS}=\sum_{f=\ell,d,u} \xi_f \frac{m_f}{v}\bar{f}f S \ . 
\end{equation}  
 Then, using the expressions for the mixing parameters in Eq.~(\ref{deltas}), we find that the couplings of the scalar $S$ to fermions are
\begin{eqnarray}
\xi_{\ell,d} &=& -\frac{2vA s^2_\beta}{m_h^2}\tan\beta \left[\frac{m_h^2}{2m_H^2}+\cot^2\beta \left(1-\frac{m_h^2}{2m_H^2} \right)  \right] \\
\xi_u &=& -\frac{2vA s^2_\beta}{m_h^2} \cot\beta \left[1-\frac{m_h^2}{2m_H^2} \left(1-\cot^2\beta \right) \right],
\end{eqnarray}
where the couplings to down-type quarks and leptons are enhanced by $\tan \beta$ and the couplings to up-type quarks are suppressed by  $\cot \beta$. In the limit of large $\tan \beta$, we may take $\beta \to \pi/2$ and $\alpha \to 0$ so that $s_\beta \to 1$ in the equations above, and we can write the couplings purely in terms of $\tan \beta$. 

We can find the couplings of $S$ to the weak gauge bosons by expanding the kinetic terms of the two scalar doublets. We find
\begin{equation}
-\mathcal{L}_{VVS}=\xi_V \frac{1}{v} \left( 2m_W^2 W^\dagger_\mu W^\mu + m_Z^2 Z_\mu Z^\mu \right)S \ ,
\end{equation}
where the coupling is the same for both $W$ and $Z$:
\begin{equation}
\xi_{W,Z}=c_\beta\delta_{13}+s_\beta\delta_{23}=\frac{-2vA s_\beta^3c_\beta}{m_h^2} \left(1+\cot^2\beta  \right) \ .
\end{equation}
In the large $\tan \beta$ limit, we write $\cos\beta \approx \cot\beta$ and $\sin\beta \to 1$ so that we can write this coupling in terms of $\cot\beta$ only:
\begin{equation}
\xi_{W,Z} = \frac{-2v A \cot\beta}{m_h^2}(1 + \cot^2\beta) \ .
\end{equation}

In summary, we have the following couplings in terms of $\tan\beta$:
\begin{eqnarray}
\xi_{\ell,d} &=& -\frac{2vA }{m_h^2}\tan\beta \left[\frac{m_h^2}{2m_H^2}+\cot^2\beta \left( 1-\frac{m_h^2}{2m_H^2} \right)  \right] ~, \\
\xi_u &=& -\frac{2vA }{m_h^2} \cot\beta \left[1-\frac{m_h^2}{2m_H^2}(1-\cot^2\beta) \right] ~, \\
\xi_{W,Z} &=& -\frac{2v A }{m_h^2} \cot\beta \, (1 + \cot^2\beta) \ .
\end{eqnarray}

\section{Coupling to Two Photons} 
\label{sec:di-photon}

To calculate the scalar coupling to two photons, we use expressions from Ref.~\cite{Carena:2012xa}, where the decay width for Higgs to two photons is given in terms of generic spin-1, spin-$\frac{1}{2}$, and spin-0 particles in the loop. Although the contribution to $S \to \gamma\gamma$ is dominated by the effective coupling $\kappa$ in the parameter region we are interested in, we include all other possible particles in the loop for completeness. In our case, there are only spin-1 and spin-$\frac{1}{2}$ particles in the loop, so the rate can be written as
\begin{equation}
\Gamma(S \to \gamma\gamma)=\frac{\alpha_{\text{EM}}^2m_S^3}{1024 \pi^3} \bigg|\frac{4\pi}{\alpha_{\text{EM}}}\kappa + \frac{g_{SVV}}{m_V^2}N_{c,V}Q_V^2 A_1(r_V)+\frac{2g_{Sf\bar{f}}}{m_f}N_{c,f}Q_f^2 A_{1/2}(r_f) \bigg|^2 \, ,
\end{equation}
where $r_i = 4m_i^2 / m_S^2$. $V$ and $f$ represent spin-1 and spin-$\frac{1}{2}$ particles, respectively; $Q$ and $N_c$ are the particle's electric charge and number of colors; and the expressions for $A_1$ and $A_{1/2}$ are given in Ref.~\cite{Carena:2012xa}.

\bibliography{lighthiggsanomalies}

\end{document}